\pdfoutput=1

\documentclass[12pt,a4paper]{article}

\usepackage{ifthen} 
\newboolean{pdflatex}
\setboolean{pdflatex}{true} 

\newboolean{articletitles}
\setboolean{articletitles}{true} 

\newboolean{uprightparticles}
\setboolean{uprightparticles}{false} 

\newboolean{inbibliography}
\setboolean{inbibliography}{false} 

\usepackage{microtype}
\usepackage{lineno}  
\usepackage{xspace} 
\usepackage{caption} 

\usepackage{graphicx}  
\usepackage{color}
\usepackage{colortbl}
\graphicspath{{./figs/}} 

\usepackage{amsmath} 
\usepackage{amssymb}
\usepackage{amsfonts}
\usepackage{upgreek} 

\newcommand*\patchAmsMathEnvironmentForLineno[1]{%
\expandafter\let\csname old#1\expandafter\endcsname\csname #1\endcsname
\expandafter\let\csname oldend#1\expandafter\endcsname\csname
end#1\endcsname
 \renewenvironment{#1}%
   {\linenomath\csname old#1\endcsname}%
   {\csname oldend#1\endcsname\endlinenomath}%
}
\newcommand*\patchBothAmsMathEnvironmentsForLineno[1]{%
  \patchAmsMathEnvironmentForLineno{#1}%
  \patchAmsMathEnvironmentForLineno{#1*}%
}
\AtBeginDocument{%
\patchBothAmsMathEnvironmentsForLineno{equation}%
\patchBothAmsMathEnvironmentsForLineno{align}%
\patchBothAmsMathEnvironmentsForLineno{flalign}%
\patchBothAmsMathEnvironmentsForLineno{alignat}%
\patchBothAmsMathEnvironmentsForLineno{gather}%
\patchBothAmsMathEnvironmentsForLineno{multline}%
\patchBothAmsMathEnvironmentsForLineno{eqnarray}%
}

\usepackage{hyperref}    
\usepackage[all]{hypcap} 

\def\lhcb {\mbox{LHCb}\xspace}

\def\velo   {VELO\xspace}

\def\MagUp {\mbox{\em Mag\kern -0.05em Up}\xspace}

\ifthenelse{\boolean{uprightparticles}}%
{

 \def\PDelta      {\ensuremath{\Delta}\xspace}                 
 \def\PXi      {\ensuremath{\Xi}\xspace}                 
 \def\PLambda      {\ensuremath{\Lambda}\xspace}                 
 \def\PSigma      {\ensuremath{\Sigma}\xspace}                 
 \def\POmega      {\ensuremath{\Omega}\xspace}                 
 \def\PUpsilon      {\ensuremath{\Upsilon}\xspace}                 
 
 \def\PB      {\ensuremath{\mathrm{B}}\xspace}                 
                  
 \def\PD      {\ensuremath{\mathrm{D}}\xspace}

 \def\PK      {\ensuremath{\mathrm{K}}\xspace}

 \def\PW      {\ensuremath{\mathrm{W}}\xspace}

 \def\PZ      {\ensuremath{\mathrm{Z}}\xspace}                 
                  
 \def\Pb      {\ensuremath{\mathrm{b}}\xspace}                 
 \def\Pc      {\ensuremath{\mathrm{c}}\xspace}

 \def\Pi      {\ensuremath{\mathrm{i}}\xspace}

}
{

 \mathchardef\PDelta="7101
 \mathchardef\PXi="7104
 \mathchardef\PLambda="7103
 \mathchardef\PSigma="7106
 \mathchardef\POmega="710A
 \mathchardef\PUpsilon="7107
                  
 \def\PB      {\ensuremath{B}\xspace}                 
                  
 \def\PD      {\ensuremath{D}\xspace}

 \def\PK      {\ensuremath{K}\xspace}

 \def\PW      {\ensuremath{W}\xspace}

 \def\PZ      {\ensuremath{Z}\xspace}                 
                  
 \def\Pb      {\ensuremath{b}\xspace}                 
 \def\Pc      {\ensuremath{c}\xspace}

 \def\Pi      {\ensuremath{i}\xspace}

}

\makeatletter
\ifcase \@ptsize \relax
  \newcommand{\miniscule}{\@setfontsize\miniscule{4}{5}}
\or
  \newcommand{\miniscule}{\@setfontsize\miniscule{5}{6}}
\or
  \newcommand{\miniscule}{\@setfontsize\miniscule{5}{6}}
\fi
\makeatother

\DeclareRobustCommand{\optbar}[1]{\shortstack{{\miniscule (\rule[.5ex]{1.25em}{.18mm})}
  \\ [-.7ex] $#1$}}

\def\W      {{\ensuremath{\PW}}\xspace}

\def\Z      {{\ensuremath{\PZ}}\xspace}

\def\cquark    {{\ensuremath{\Pc}}\xspace}

\def\bquark    {{\ensuremath{\Pb}}\xspace}

  \def\Kbar    {{\kern 0.2em\overline{\kern -0.2em \PK}{}}\xspace}

\def\KorKbar    {\kern 0.18em\optbar{\kern -0.18em K}{}\xspace}

  \def\Dbar    {{\kern 0.2em\overline{\kern -0.2em \PD}{}}\xspace}

\def\DorDbar    {\kern 0.18em\optbar{\kern -0.18em D}{}\xspace}

\def\Bbar    {{\ensuremath{\kern 0.18em\overline{\kern -0.18em \PB}{}}}\xspace}

\def\BorBbar    {\kern 0.18em\optbar{\kern -0.18em B}{}\xspace}

  \def\Y#1S{\ensuremath{\PUpsilon{(#1S)}}\xspace}

\def\Lz          {{\ensuremath{\PLambda}}\xspace}
\def\Lbar        {{\ensuremath{\kern 0.1em\overline{\kern -0.1em\PLambda}}}\xspace}
\def\LorLbar    {\kern 0.18em\optbar{\kern -0.18em \PLambda}{}\xspace}

\def\to                 {\ensuremath{\rightarrow}\xspace}

\def\AT#1     {\ensuremath{A_{\mathrm{T}}^{#1}}\xspace}           

\def\C#1      {\ensuremath{\mathcal{C}_{#1}}\xspace}                       
\def\Cp#1     {\ensuremath{\mathcal{C}_{#1}^{'}}\xspace}                    
\def\Ceff#1   {\ensuremath{\mathcal{C}_{#1}^{\mathrm{(eff)}}}\xspace}        
\def\Cpeff#1  {\ensuremath{\mathcal{C}_{#1}^{'\mathrm{(eff)}}}\xspace}       
\def\Ope#1    {\ensuremath{\mathcal{O}_{#1}}\xspace}                       
\def\Opep#1   {\ensuremath{\mathcal{O}_{#1}^{'}}\xspace}                    

\newcommand{\tev}{\ifthenelse{\boolean{inbibliography}}{\ensuremath{~T\kern -0.05em eV}\xspace}{\ensuremath{\mathrm{\,Te\kern -0.1em V}}}\xspace}
\newcommand{\gev}{\ensuremath{\mathrm{\,Ge\kern -0.1em V}}\xspace}
\newcommand{\mev}{\ensuremath{\mathrm{\,Me\kern -0.1em V}}\xspace}
\newcommand{\kev}{\ensuremath{\mathrm{\,ke\kern -0.1em V}}\xspace}
\newcommand{\ev}{\ensuremath{\mathrm{\,e\kern -0.1em V}}\xspace}
\newcommand{\gevc}{\ensuremath{{\mathrm{\,Ge\kern -0.1em V\!/}c}}\xspace}
\newcommand{\mevc}{\ensuremath{{\mathrm{\,Me\kern -0.1em V\!/}c}}\xspace}
\newcommand{\gevcc}{\ensuremath{{\mathrm{\,Ge\kern -0.1em V\!/}c^2}}\xspace}
\newcommand{\gevgevcccc}{\ensuremath{{\mathrm{\,Ge\kern -0.1em V^2\!/}c^4}}\xspace}
\newcommand{\mevcc}{\ensuremath{{\mathrm{\,Me\kern -0.1em V\!/}c^2}}\xspace}

\def\mm   {\ensuremath{\rm \,mm}\xspace}

\def\mum  {\ensuremath{{\,\upmu\rm m}}\xspace}

\def\invfb   {\ensuremath{\mbox{\,fb}^{-1}}\xspace}

\def\gsim{{~\raise.15em\hbox{$>$}\kern-.85em
          \lower.35em\hbox{$\sim$}~}\xspace}
\def\lsim{{~\raise.15em\hbox{$<$}\kern-.85em
          \lower.35em\hbox{$\sim$}~}\xspace}

\def\sqs   {\ensuremath{\protect\sqrt{s}}\xspace}

\def\ptot       {\mbox{$p$}\xspace}
\def\pt         {\mbox{$p_{\rm T}$}\xspace}

\def\mrad{\ensuremath{\rm \,mrad}\xspace}
\def\rad{\ensuremath{\rm \,rad}\xspace}

\newcommand{\lum} {\ensuremath{\mathcal{L}}\xspace}

\def\geant      {\mbox{\textsc{Geant4}}\xspace}

\def\pythia     {\mbox{\textsc{Pythia}}\xspace}

\def\tell1  {TELL1\xspace}
\def\ukl1   {UKL1\xspace}

\newcommand{\eg}{\mbox{\itshape e.g.}\xspace}

\usepackage{cite} 
\usepackage{mciteplus}

\usepackage{longtable} 

\newcommand{\tevc}{\ensuremath{{\mathrm{\,Te\kern -0.1em V\!/}c}}\xspace}

\begin{document}

\renewcommand{\thefootnote}{\fnsymbol{footnote}}
\setcounter{footnote}{1}


\begin{titlepage}
\pagenumbering{roman}

\vspace*{-1.5cm}
\centerline{\large EUROPEAN ORGANIZATION FOR NUCLEAR RESEARCH (CERN)}
\vspace*{1.5cm}
\hspace*{-0.5cm}
\begin{tabular*}{\linewidth}{lc@{\extracolsep{\fill}}r}
\ifthenelse{\boolean{pdflatex}}
{\vspace*{-2.7cm}\mbox{\!\!\!\includegraphics[width=.14\textwidth]{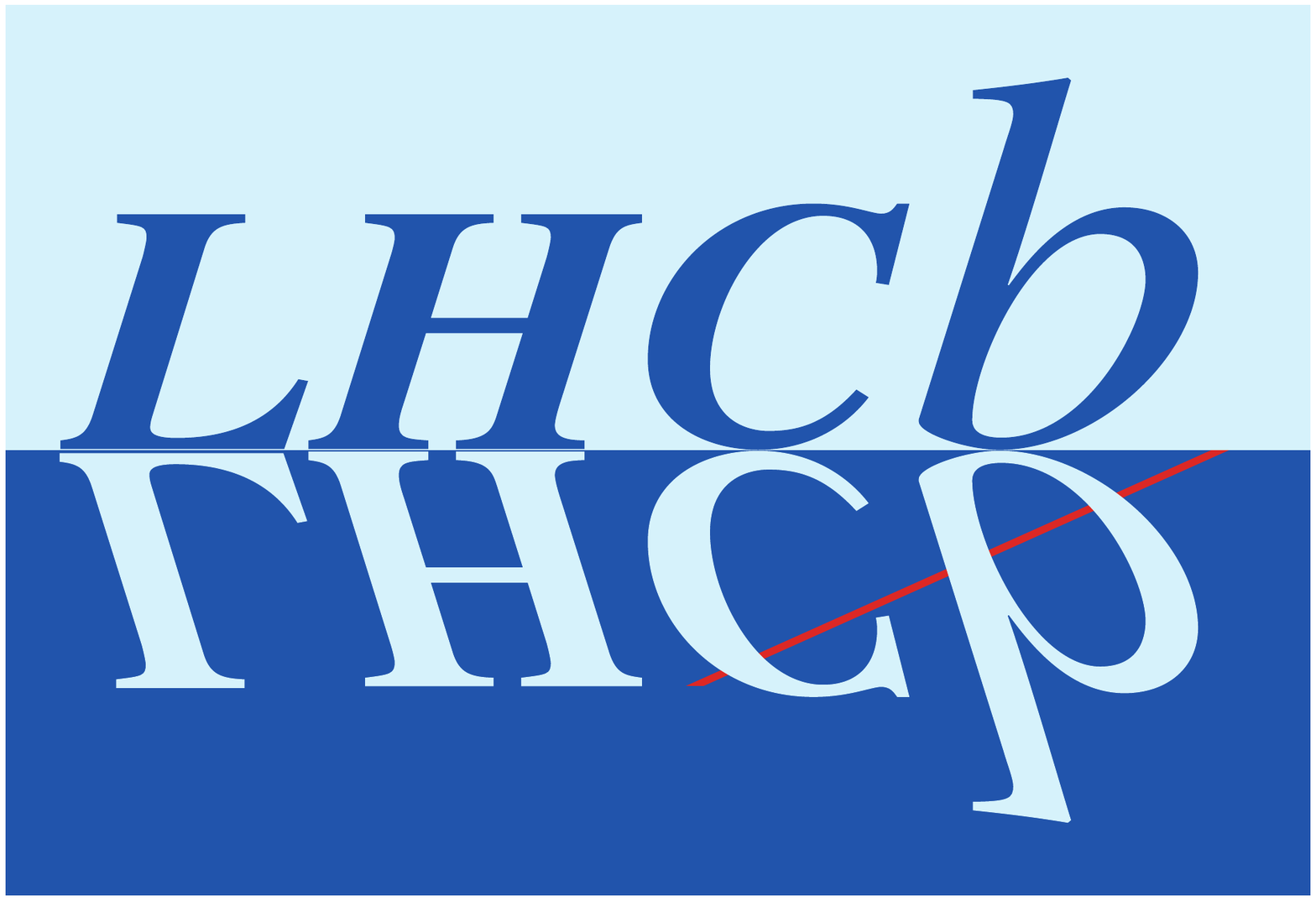}} & &}%
{\vspace*{-1.2cm}\mbox{\!\!\!\includegraphics[width=.12\textwidth]{lhcb-logo.eps}} & &}%
\\
 & & CERN-PH-EP-2105-139 \\  
 & & LHCb-PAPER-2015-002 \\  
 & & 28 September 2015 \\
 & & \\
\end{tabular*}

\vspace*{2.0cm}

{\bf\boldmath\huge
\begin{center}
  Search for  long-lived heavy charged particles using a
  ring imaging Cherenkov technique at LHCb
\end{center}
}

\vspace*{2.0cm}

\begin{center}
The LHCb collaboration\footnote{Authors are listed at the end of this paper.}
\end{center}

\vspace{\fill}

\begin{abstract}
\noindent
A search is performed for heavy long-lived charged particles
using 3.0\invfb of proton-proton collisions collected at \sqs= 7 and 8\tev with the LHCb detector.
The search is mainly based on the response of the ring imaging Cherenkov
detectors to distinguish
the heavy, slow-moving particles from muons.
No evidence is found for the production of such long-lived states. 
The results are expressed as limits on the Drell-Yan production of pairs
of long-lived particles, with both particles in the LHCb pseudorapidity
acceptance, $1.8 < \eta < 4.9$.
The mass-dependent cross-section upper limits are in
the range 2--4~fb (at 95\% CL)
for masses between 124 and 309\gevcc.

\end{abstract}

\vspace*{2.0cm}

\begin{center}
  Published in Eur. Phys. J. C 75 (2015) 595
\end{center}

\vspace{\fill}

{\footnotesize 
\centerline{\copyright~CERN on behalf of the \lhcb collaboration, licence \href{http://creativecommons.org/licenses/by/4.0/}{CC-BY-4.0}.}}
\vspace*{2mm}

\end{titlepage}


\newpage
\setcounter{page}{2}
\mbox{~}

\cleardoublepage


\renewcommand{\thefootnote}{\arabic{footnote}}
\setcounter{footnote}{0}



\pagestyle{plain} 
\setcounter{page}{1}
\pagenumbering{arabic}


\newcommand{\lumi}{3\xspace}
\newcommand{\lumia}{1.01\xspace}
\newcommand{\lumib}{1.98\xspace}
\newcommand{\stau}{\ensuremath{\widetilde{\tau_1}}\xspace}

%
%
\section{Introduction}

Several extensions of the Standard Model (SM) propose the existence of
charged massive stable particles (CMSP).
Stable particles, in this context, are long-lived particles that can travel through a detector
without decaying.
These particles can have long lifetimes for a variety of reasons,
\eg a new (approximately) conserved quantum number, a weak coupling or a
limited phase space in any allowed decay.
In supersymmetric (SUSY) models, CMSPs
can be sleptons ($\widetilde{\ell}$), charginos,
or R-hadrons.
R-hadrons are colourless states combining squarks ($\widetilde{q}$) or gluinos ($\widetilde{g}$) and SM quarks
or gluons.
In the gauge-mediated supersymmetry breaking (GMSB) model~\cite{Dimopoulos,gmsb_giudice, susy_martin}
the breakdown of SUSY is mediated by gauge interactions and can occur at a relatively low
energy scale. For a particular range of parameter space in the minimal model (mGMSB)
the next-to-lightest supersymmetric particle
can be a long-lived stau (\stau),
with a mass of the order of 100\gevcc or higher.
The \stau is the lightest mass eigenstate, resulting from the mixture
of right-handed and left-handed superpartners of the $\tau$, dominated by the right-handed component.

A CMSP loses energy mainly via ionisation;
strongly interacting CMSPs are not considered here.
In a detector such as LHCb a CMSP with a kinetic energy above about 5\gev
should be able to traverse the muon chambers.
Those particles would often be produced with a relatively low velocity
and could be identified by their time-of-flight, and by
their specific energy loss, $\mathrm{d}E/\mathrm{d}x$, in the detectors;
Cherenkov radiation would be absent in Cherenkov counters tuned for ultra-relativistic particles.

Several experiments have searched for CMSPs~\cite{ALEPH,DELPHI,L3,OPAL,H1,limit_D0,limit_CDF,limit_atlas_n,limit_cms}.
With the exception of DELPHI~\cite{DELPHI}, which had Cherenkov counters,
the analyses are based on   $\mathrm{d}E/\mathrm{d}x$ and time-of-flight measurements.
The primary interest here is to show the potential of the identification technique
based on  ring imaging Cherenkov (RICH) detectors,
in addition to the exploration of the forward
pseudorapidity region only partially covered by the central detectors
at the Tevatron and the LHC.

The analysis described in this study is
mainly based on the absence of Cherenkov radiation in the RICH detectors.
This technique is used to search for pairs of CMSPs in LHCb, produced by
a Drell-Yan mechanism.

%
%
%
%
%
%

\section{The \lhcb detector and the detection of slow particles}
\label{detector}

The \lhcb detector~\cite{Alves:2008zz,LHCb-DP-2014-002} is a single-arm forward
spectrometer covering the approximate \mbox{pseudorapidity} range $1.8<\eta <4.9$,
designed for the study of particles containing \bquark or \cquark
quarks.
The detector includes a high-precision tracking system
consisting of a silicon-strip vertex detector (the vertex locator, \velo)
surrounding the proton-proton interaction region~\cite{LHCb-DP-2014-001},
a large-area silicon-strip detector located
upstream of a dipole magnet with a bending power of about
$4{\rm\,Tm}$, and three stations of silicon-strip detectors and straw
drift tubes~\cite{LHCb-DP-2013-003} placed downstream of the magnet.
The tracking system provides a measurement of momentum, \ptot, of charged particles with
a relative uncertainty that varies from 0.5\% at low momentum to 1.0\% at 200\gevc.
The minimum distance of a track to a primary vertex (PV), the impact parameter (IP),
is measured with a resolution of $(15+29/\pt)\mum$,
where \pt is the component of the momentum transverse to the beam, in\,\gevc.
Photons, electrons and hadrons are identified by a calorimeter system consisting of
scintillating-pad and preshower detectors, an electromagnetic
calorimeter and a hadronic calorimeter. Muons are identified by a
system composed of alternating layers of iron and multiwire
proportional chambers~\cite{LHCb-DP-2012-002}.

Different types of charged particles are distinguished using information from two RICH
detectors~\cite{LHCb-DP-2012-003}.
The RICH system, which plays a crucial role in this analysis, consists of an upstream detector
with silica aerogel and $\rm C_4F_{10}$ gas radiators, positioned directly after the \velo,
and a downstream detector with a $\rm CF_4$ gas radiator,
located just after the tracking system.

The online event selection is performed by a trigger~\cite{LHCb-DP-2012-004}, 
which consists of a hardware stage, based on information from the calorimeter and muon
systems, followed by a software stage, which applies a full event
reconstruction.

The analysis presented here is based on two data sets collected in 2011 and 2012
corresponding to integrated luminosities of  1.0\invfb and 2.0\invfb from
proton-proton collisions recorded
at centre-of-mass energies of 7 and 8\tev, respectively.

\begin{figure}[tb]
\begin{center}
  \includegraphics[width=0.55\linewidth]{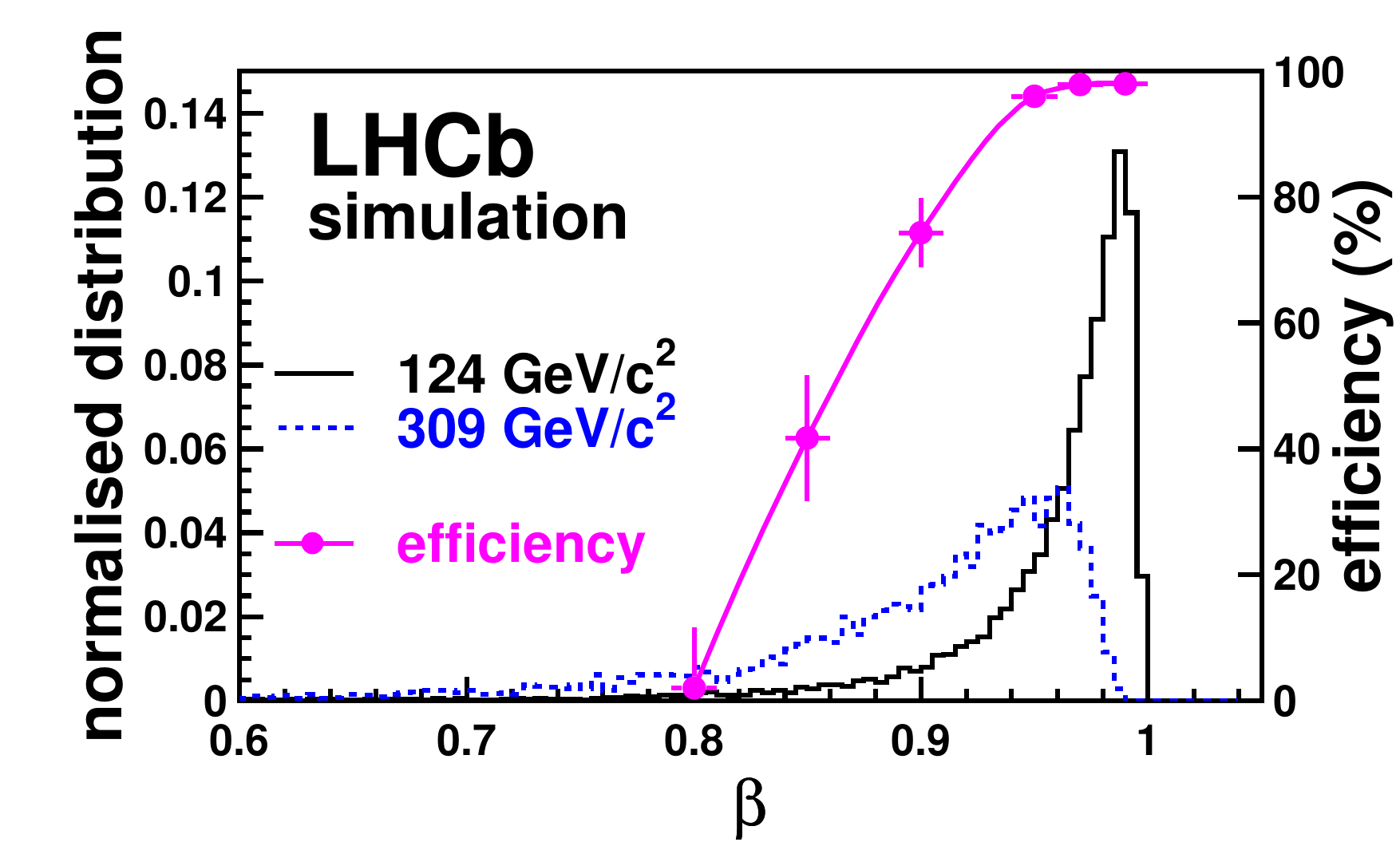}
\caption{\small
  CMSP velocity spectrum for the CMSP masses of 124\gevcc and 309\gevcc. The proton-proton
  centre-of-mass energy is 7~TeV.
  The dots with error bars show the efficiency to detect tracks as a function of the $\beta$
  of the particle (right scale).}
\label{fig:beta}
\end{center}
\end{figure}

In the production process considered, CMSPs can have velocities $\beta \equiv v/c$
as low as 0.7, and their arrival time at the subdetectors
can differ by several nanoseconds with respect to lighter particles with $\beta \simeq 1$.
For illustration, the $\beta$ spectrum is shown in Fig.~\ref{fig:beta}, for two values
of the CMSP mass, at centre-of-mass energy of 7\tev.
The effects of such delayed detection on the efficiencies of the subdetectors are determined from
simulation in which the timing information is modelled according to dedicated electronic
measurements and tests in beam. 
The muon chambers have the largest inefficiency for slow-particle reconstruction.
The maximal delay for a particle to be accepted by the front-end electronics is 12.5~ns~\cite{LHCb-DP-2012-002}.
In the most downstream muon chamber, this delay corresponds to the arrival of
a particle with $\beta=0.83$.
To be identified as a muon, the charged particle must be associated with hits in the
last four muon chambers,
a requirement that particles with $\beta < 0.8$ fail to meet. 
The large time-of-flight can also bias the reconstructed position of the particle passing through the
tracker straw tubes, which accept a maximal drift-time of about 35~ns for tracks passing close to
the straw radius of 2.5~mm~\cite{LHCb-DP-2013-003}.
These combined effects result in a vanishingly small reconstruction efficiency
for particles with $\beta < 0.8$ but an efficiency above 95\% if $\beta > 0.95$, as shown 
in Fig.~\ref{fig:beta}.

%
%
\section{Simulation}
\label{MC}
%

\subsection{CMSP signal}
\label{sec:gen-stau}
The adopted framework is stau pair production, $\stau^+ \stau^-$, in mGMSB via a Drell-Yan process.
Pairs of CMSPs originating from cascade decays of heavier particles are explicitly not
considered.
In the following the symbol \stau is used when the context is explicitly the mGMSB model,
while CMSP is kept for the more general context.

The mGMSB model has six parameters~\cite{susy_martin,gmsb_giudice}: the SUSY breaking scale ($\Lambda$), the
mass scale of the SUSY loop messengers ($M_m$), the number of messenger supermultiplets ($N_5$), the ratio
of the vacuum expectation values of the two neutral Higgs fields
($\tan \beta$), the sign of the Higgs mass parameter ($\mu$), and the parameter
$C_{\text{grav}}$, which affects the gravitino mass.
The \textsc{Spheno3.0} SUSY spectrum generator~\cite{spheno} is used
to compute the masses of the \stau as a function of the above six parameters.
The SPS7 benchmark scenario~\cite{SPS_benchmarks}
is used to determine the parameter space, where $N_5 = 3$, $\tan\beta= 15$, $\mu > 0$,
$M_m = 2\Lambda$, and the parameter $C_{\text{grav}} = 4000$ are fixed.
Variation of $\Lambda$ then uniquely determines the \stau mass and lifetime, which is of the order
of 100 ns. In this study the \stau is considered stable.\par

\begin{table}[tb]
\begin{center}
  \caption{\small Values of the mGMSB $\Lambda$ parameters in the SPS7 scenario used in this study,
  the corresponding masses of the \stau, $m_{\widetilde{\tau}}$, and the
  cross-section of the pair production at next-to-leading order.
  The last two columns give the detector acceptance  $A$.
  }
  \label{tab:gmsb1}
  \begin{tabular}{  c  c  c c   c c }
  \hline $\Lambda$ & $m_{\widetilde{\tau}}$ & \multicolumn{2}{c }{$\sigma$ (fb)} &  \multicolumn{2}{c }{$A$ (\%)}  \\
    (\tev) &(GeV/c$^2$) & 7\tev & 8\tev & 7\tev & 8\tev \\

  \hline
  40 & 124  &   16.90 $\pm$ 0.79 & 21.20   $\pm$ 0.91&  8.3  &    9.5 \\
  50 & 154  & \, 7.19 $\pm$ 0.38 & \, 9.20 $\pm$ 0.46&  6.5  &    7.7 \\
  60 & 185  & \, 3.44 $\pm$ 0.20 & \, 4.50 $\pm$ 0.24&  5.2  &    6.1 \\
  70 & 216  & \, 1.79 $\pm$ 0.11 & \, 2.39 $\pm$ 0.14&  4.3  &    5.0 \\
  80 & 247  & \, 1.00 $\pm$ 0.07 & \, 1.35 $\pm$ 0.08&  3.4  &    4.1 \\
  90 & 278  & \, 0.57 $\pm$ 0.04 & \, 0.80 $\pm$ 0.05&  2.8  &    3.4 \\
 100 & 309  & \, 0.34 $\pm$ 0.02 & \, 0.49 $\pm$ 0.03&  2.3  &    2.9 \\
  \hline
\end{tabular}
\end{center}
\end{table}

The predictions for \stau pair production are based on
next-to-leading order (NLO) cross-section calculations by the
\textsc{Prospino2.1} program~\cite{prospino_paper}
using the CTEQ6.6M parton distribution function (PDF) set~\cite{cteq66}.
These predictions at \sqs= 7 and 8\tev are presented in
Table~\ref{tab:gmsb1}. The relative theoretical uncertainties vary between 5\% and 8\%,
and are determined following Ref.~\cite{prospino-errors}.

Fully simulated signal samples,
with masses varying from 124 to 309 GeV/$c^2$, have been produced
for proton-proton collisions at \sqs= 7 and  8\tev.
The \stau pairs generated by \pythia 6.423~\cite{Sjostrand:2006za},
with both \stau particles in the fiducial range $1.8 < \eta < 4.9$ are passed to
\geant~\cite{Agostinelli:2002hh,LHCb-PROC-2011-006} for detector simulation.
The fraction of \stau pairs within the fiducial range is defined as the acceptance, $A$.
The acceptance factor obtained from \pythia
with the MSTW2008 PDF set~\cite{mstw08} is also shown in Table~\ref{tab:gmsb1},
with model uncertainties ranging from 5\% to 9\% for
\stau  mass from 124\gevcc to 309\gevcc, mainly associated to the
choice of PDF.

For larger \stau masses,
the Drell-Yan process results in a lower forward boost of the \stau pair,
with a subsequent increase in the pair opening angle in the detector frame.
The decrease of $A$ for an increasing \stau mass is due to a higher probability
for one of the particles to escape the LHCb geometrical acceptance.

\subsection{Background}
\label{bgr}
The main background is from the  Drell-Yan production of muon pairs,
$\Z/\gamma^{\star} \rightarrow \mu^+ \mu^-$.
Samples of $Z/\gamma^{\star} \rightarrow \mu^+ \mu^-$ events have been produced with
\pythia and fully simulated with \geant.  The cross-section for this process
has been calculated with \textsc{DYNNLO}~\cite{dynnlo} at next-to-next-to-leading order
with the MSTW2008 PDF set.
The preselection requirements (see Sect.~\ref{pre_sel})
lead to values of the predicted cross-section in LHCb for
\sqs=7 and 8\tev of $1.08 \pm 0.10$ and $1.36 \pm 0.12$~pb, respectively.
These values are nearly two orders of magnitude larger than
the predicted \stau pair cross-section in the most favourable case,
corresponding to $\Lambda=40$\tev.

Other background sources include muons produced
by top quark pairs, and from $\tau$ pairs.
To study the background contributions from these processes, samples of
$\Z/\gamma^{\star} \rightarrow \tau^+ \tau^-$ and
top quark pair decays have been simulated.

%
%
%
%
%

\section{Data selection} 
The event selection is performed in two steps:
a preselection
aimed at suppressing the most prominent backgrounds,
followed by a multivariate analysis, based on an artificial neural network
that is trained using calibrated simulation.

\subsection{Preselection}
\label{pre_sel}

CMSP candidates are identified as high-momentum charged particles with hits
in the \velo, all the tracking stations and the four last
muon detectors. 

Events are selected that contain two or more such particles where
one of the particles
passes the high-\pt single muon trigger with a threshold of 15\gevc.
The trigger efficiency is estimated from simulation to be 92\% for
a mass of 124\gevcc, and 89\% for 309\gevcc.
The two candidates must have opposite charge and each have $\pt>50$\gevc.
To reject background from $Z/\gamma^{\star} \rightarrow \mu^+ \mu^-$
the pair must have a dimuon mass larger than 100\gevcc.
A mass-dependent lower threshold on momentum
is applied to select particles with $\beta > 0.8$.

Several criteria are used to reject muons,
electrons and hadrons.
Pions and kaons in jets may be identified as muons if they decay in flight or if shower fragments
escape from the calorimeters to the muon stations.
As hadrons and electrons deposit more energy in the calorimeters than that expected for CMSPs,
an efficient rejection of these backgrounds is achieved
by requiring the sum of the ECAL and HCAL energies associated with the
extrapolation of the charged particle to the calorimeters
to be less than 1\% of the momentum of that particle.
The background from misidentified muons contributes approximately equally
to same- and opposite-charge pairs~\cite{LHCb-PAPER-2012-008}.
No same-charge event is found in the preselected data, showing that this contribution is negligible.

CMSPs, as well as muons from $\Z/\gamma^{\star}$ decays,
would be  produced at the PV and should have a smaller IP with respect to the PV than
muons from heavy quark or tau decays.
Requiring an IP of less than 50 $\mu$m selects efficiently CMSP candidates.
After preselection, the contribution from the
$\Z/\gamma^{\star} \rightarrow \tau^+ \tau^-$
process where both taus decay leptonically
to muons is estimated from simulation to contribute less than 0.1 events in total.
Pairs of muons produced from top quark decays into $b$ quarks and $W^\pm$
bosons, with the  $W^\pm$ bosons decaying leptonically into muons,
contribute less than one event, as determined from simulation.

In summary, after preselection the only significant source of background is from  $\Z/\gamma^{\star} \rightarrow \mu^+ \mu^-$
decays.
The predicted number of dimuon events in the 7\tev (8\tev) data set
is 249$\pm$49 (570$\pm$110) which is in good agreement with the 239 (713) observed
candidate events.
The uncertainties comprise contributions from the preselection cuts (Section~\ref{results}),
and the uncertainty on the $\Z/\gamma^{\star} \rightarrow \mu^+ \mu^-$ cross-section (Section~\ref{bgr}).
The expected number of events with \stau pairs is
2.7 events in the full data set of \lum = 3.0\invfb,
according to the cross-section calculated with \textsc{Prospino2.1},
with SPS7 parameters and a \stau mass of 124\gevcc.

\subsection{Selection}\label{ANN_signal_det}

An artificial neural network (ANN) is used to distinguish CMSPs from muons by exploiting the
difference in interactions that these particles have in matter.
To reduce model dependence, the ANN is applied to the individual CMSP candidates,
rather than to CMSP-pairs, and a minimum requirement is placed on the product of the two ANN responses.
Four variables of the CMSP candidates are used as ANN inputs, computed from
the energy deposited in the \velo sensors ($\Delta$E VELO), in the ECAL ($\Delta$E ECAL),
in the HCAL ($\Delta$E HCAL),
and a likelihood variable associated with the RICH information (DLLx).
Model dependence is reduced as much as possible by the absence of kinematical
observables in the ANN.
The energy loss of a charged particle traversing a \velo sensor follows a Landau distribution.
The most probable energy deposition in a sensor is estimated using a truncated mean where only the 60\% lowest
depositions are averaged.

\begin{table}[bt]
\begin{center}
\caption{\small
Number of events with both CMSPs candidates satisfying $\rm DLLx>-5$, 
and the final efficiency, $\epsilon$, after the multivariate analysis selection,
given for each mass hypothesis.}
\begin{tabular}{ c   c  c  c   c c  }
 \hline
\vspace{-1mm}
 &                         \multicolumn{2}{c}{\small CMSP-pair } & &\\
$m_{\rm CMSP}$   & \multicolumn{2}{c}{\small candidates } &  \multicolumn{2}{c}{ $\epsilon$ (\%)}   \\ 
(\gevcc) & 7\tev   &  8\tev   & 7\tev & 8\tev   \\
\hline
124  &     38 &   73 &    49.6 $\pm$     4.4 &    45.1 $\pm$     4.4 \\ 
154  &     36 &   68 &    48.9 $\pm$     4.5 &    44.5 $\pm$     4.5 \\ 
185  &     36 &   68 &    46.0 $\pm$     4.7 &    41.9 $\pm$     4.7 \\ 
216  &     28 &   56 &    42.0 $\pm$     4.8 &    38.5 $\pm$     4.8 \\ 
247  &     24 &   49 &    37.5 $\pm$     5.0 &    35.0 $\pm$     5.0 \\ 
278  &     24 &   49 &    32.8 $\pm$     5.1 &    31.2 $\pm$     5.1 \\ 
309  &     13 &   30 &    28.4 $\pm$     5.3 &    27.3 $\pm$     5.3 \\ 
\hline
\end{tabular}
\label{tab:summary}
\end{center}
\end{table}

Particle identification for a CMSP candidate, using RICH information,
is provided by the DLLx variable.
A particle identification hypothesis is assigned to a track using a likelihood method.
The information from the three radiators is combined and a ``delta log-likelihood'' (DLL)
value computed. The DLL gives, for each track, the change in the overall event log-likelihood when the
particle ID hypothesis is changed from $\pi$ to $\mu$, $e$, K, p.
The DLLx classification has been added to account for high momentum particles which do not radiate, or have
a Cherenkov angle which is too small to fit one of the five particle hypotheses.
A positive DLLx indicates a high probability that the candidate has a relatively low velocity.
More details are given in Section~\ref{sec:riches}.

Events with both candidate CMSPs with $\rm DLLx >-5$ are used in the analysis,
with no loss of signal, as deduced from simulation.
The numbers of selected events with CMSP-pairs are given in Table~\ref{tab:summary}.

\begin{figure}[tb]
\begin{center}
\includegraphics[width=0.48\linewidth]{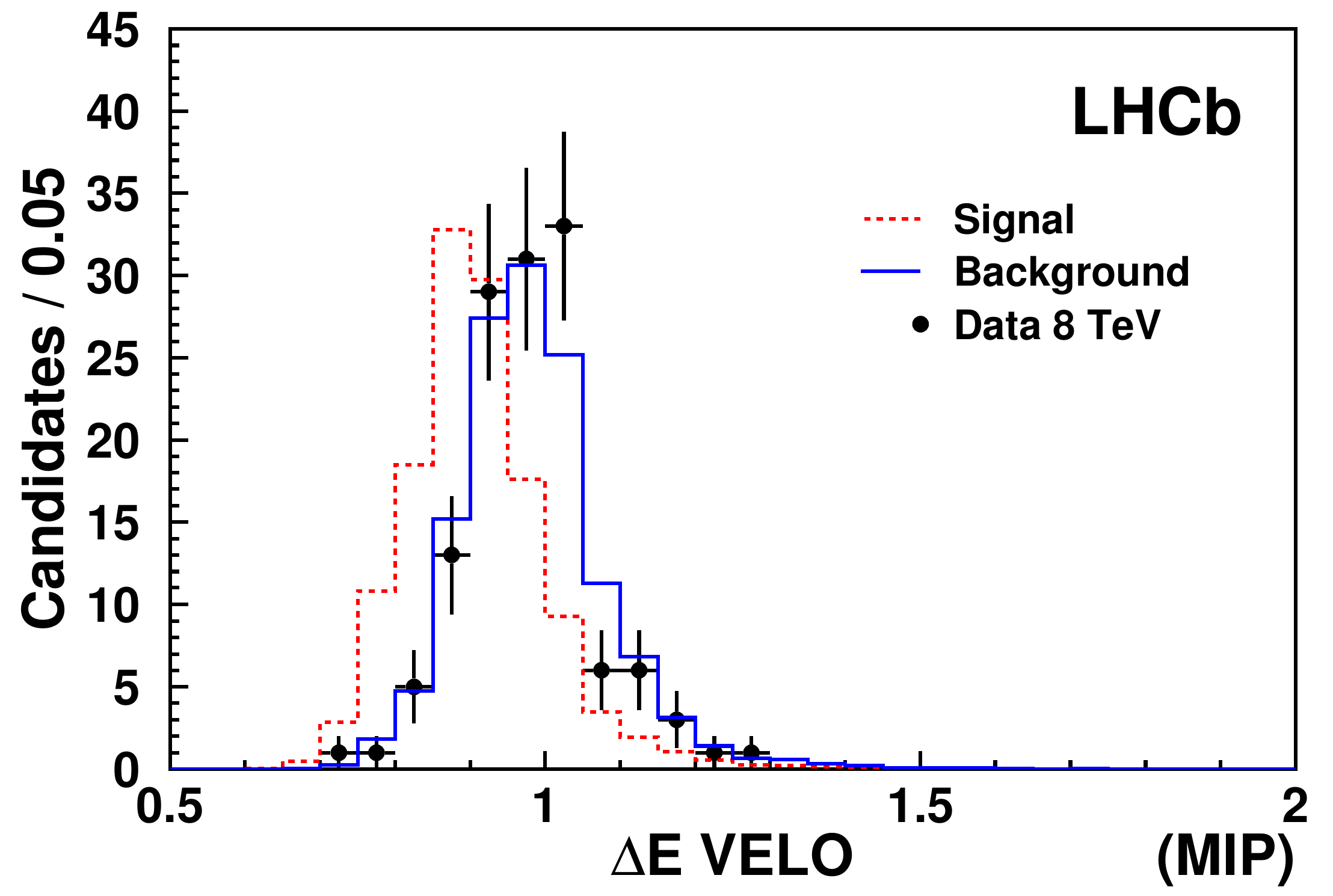}
\hspace*{1mm}\includegraphics[width=0.495\linewidth]{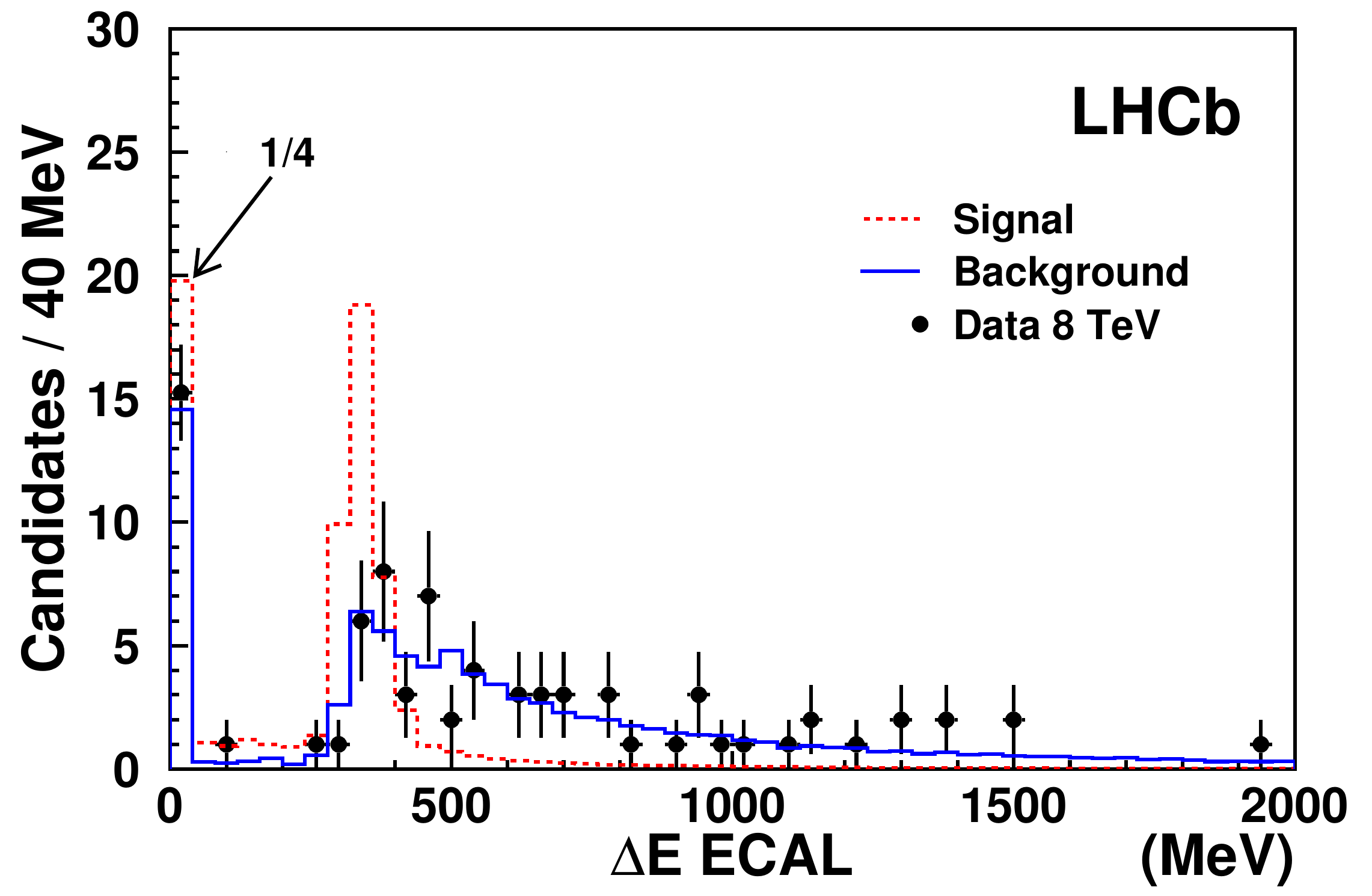} 
\includegraphics[width=0.49\linewidth]{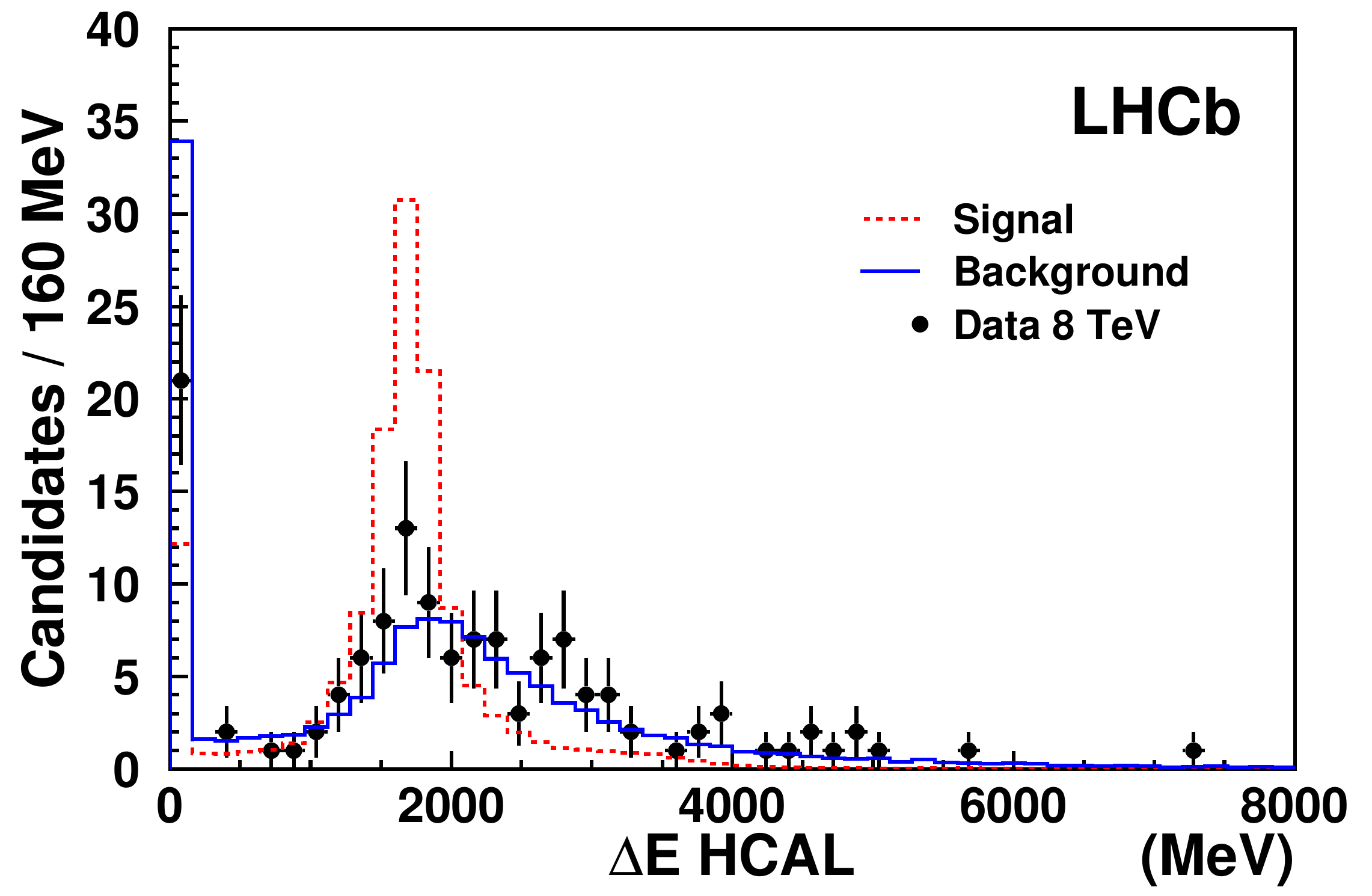}
\includegraphics[width=0.49\linewidth]{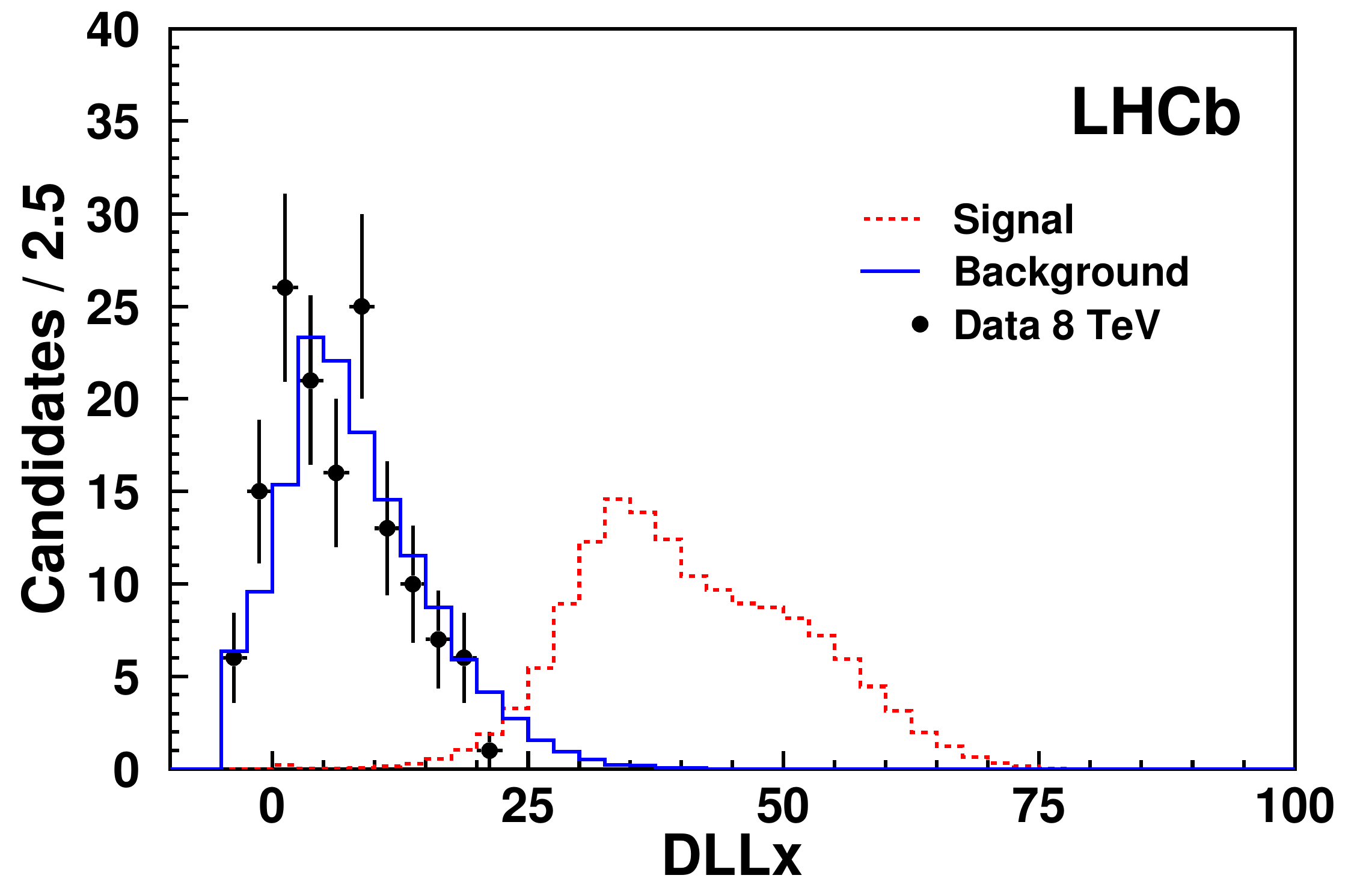} 
\caption{\small
  Number of CMSP candidates, as a function of the four variables used as inputs to the
  ANN. There are two CMSP candidates per event. The black dots with error bars show the 2012 data.
  The dashed red histogram is the expected shape for 124\gevcc
  CMSPs and the blue histogram shows the background from $\Z/\gamma^{\star}$ decays into muons.
  The energy in the \velo is given in units of minimum ionising particle (MIP) deposition.
  The first bin of the histogram for $\Delta$E in the ECAL has been multiplied by a factor 0.25.}
\label{fig:4var_forANN}
\end{center}
\end{figure}

\begin{figure}[t]
\begin{center}
\includegraphics[width=0.49\linewidth]{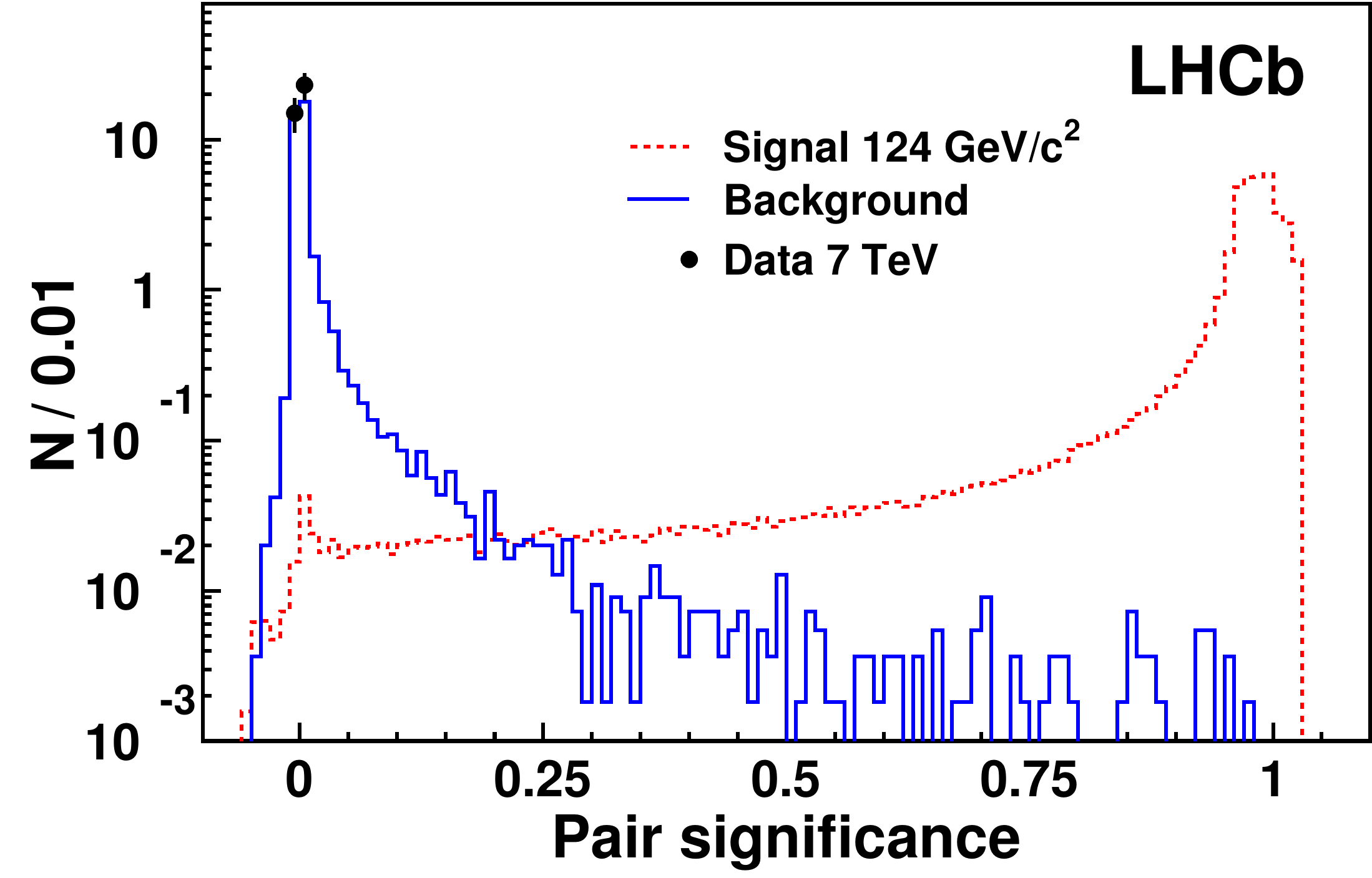}
\thicklines \put(-122,80){\vector(0,-5){60}}
\hspace*{2mm}\includegraphics[width=0.49\linewidth]{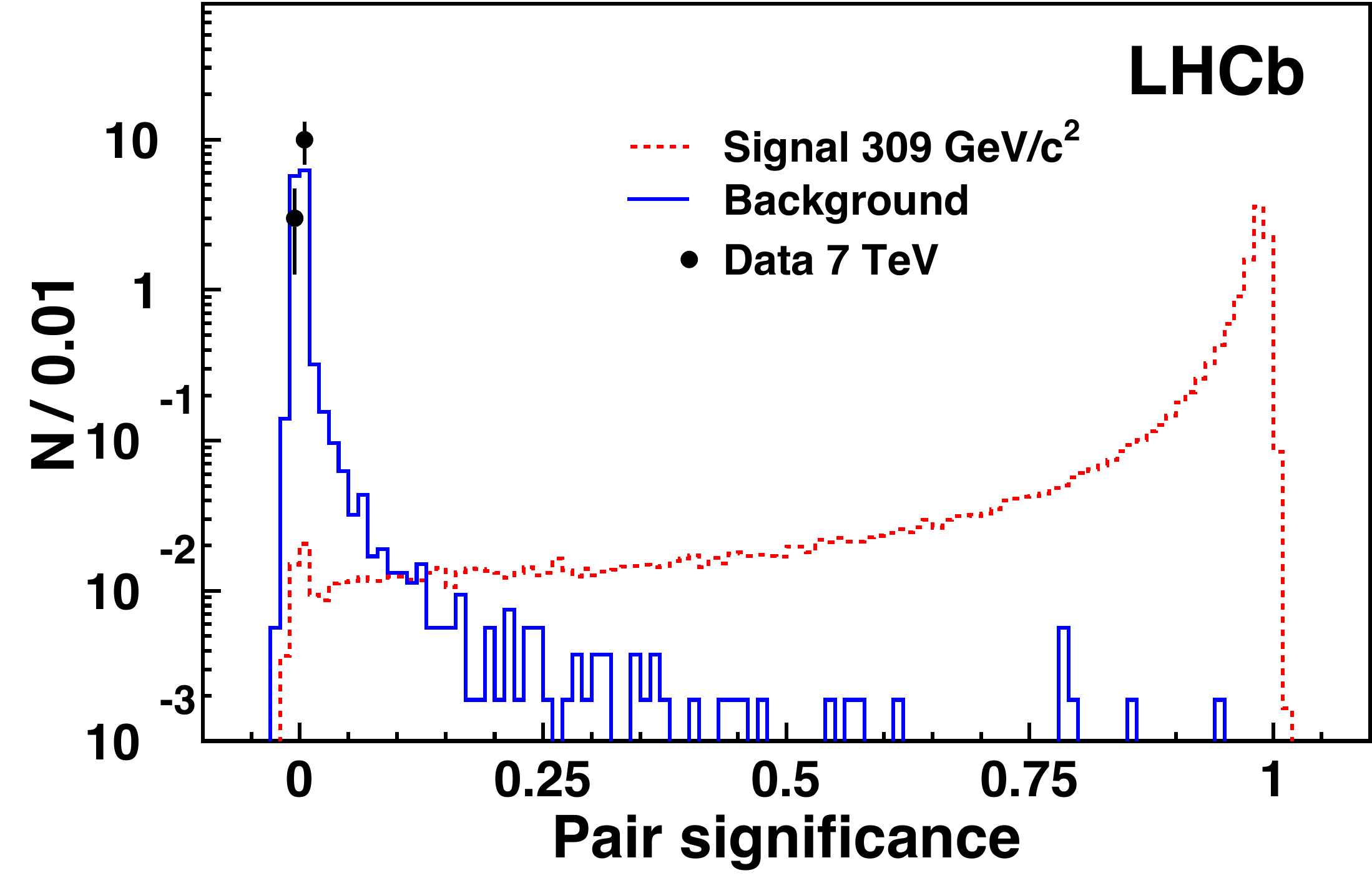}
\put(-119,80){\vector(0,-5){60}}
\vspace*{5mm}
\includegraphics[width=0.49\linewidth]{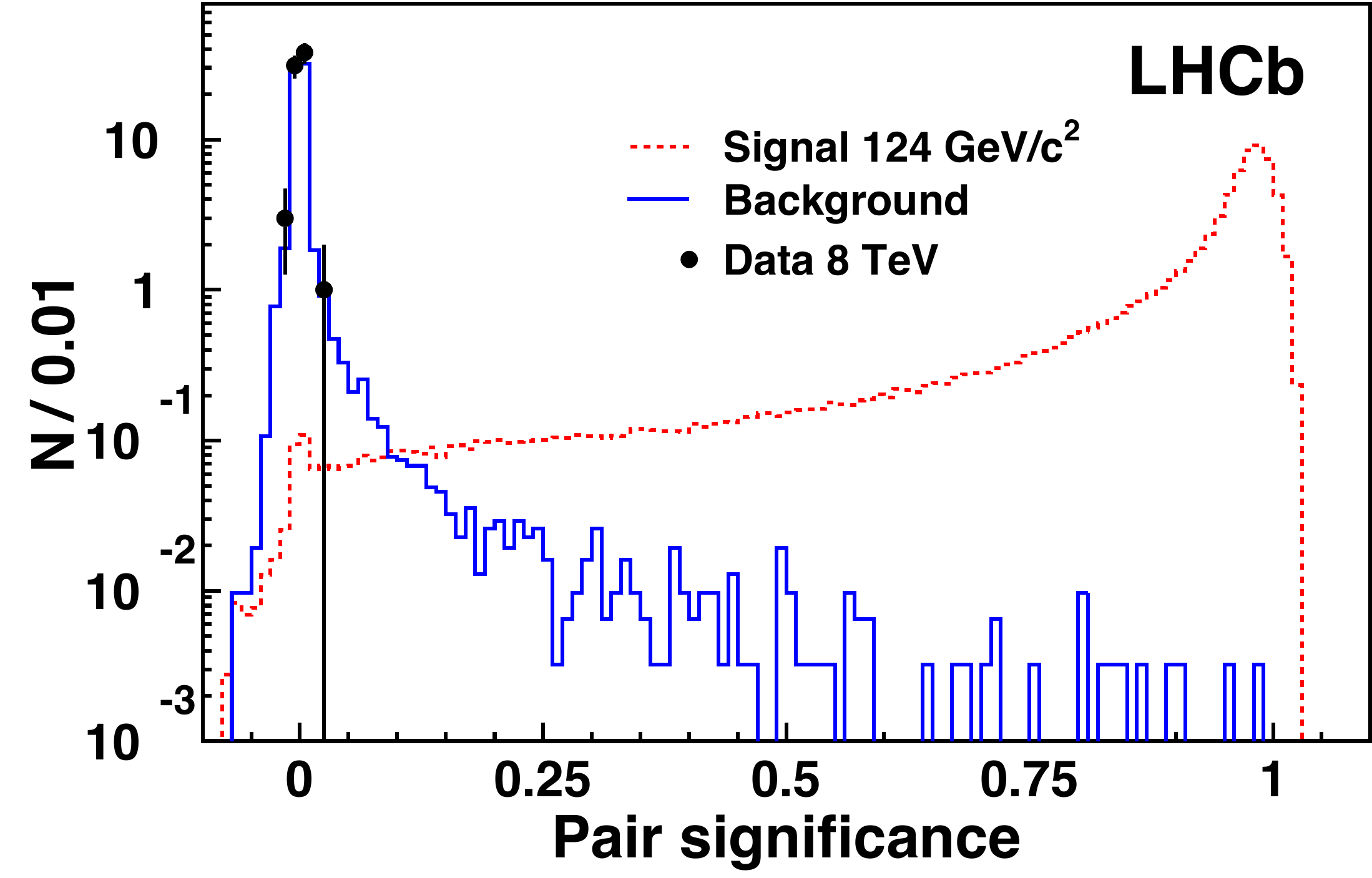}
\thicklines \put(-117,80){\vector(0,-5){60}}
\hspace*{2mm}\includegraphics[width=0.49\linewidth]{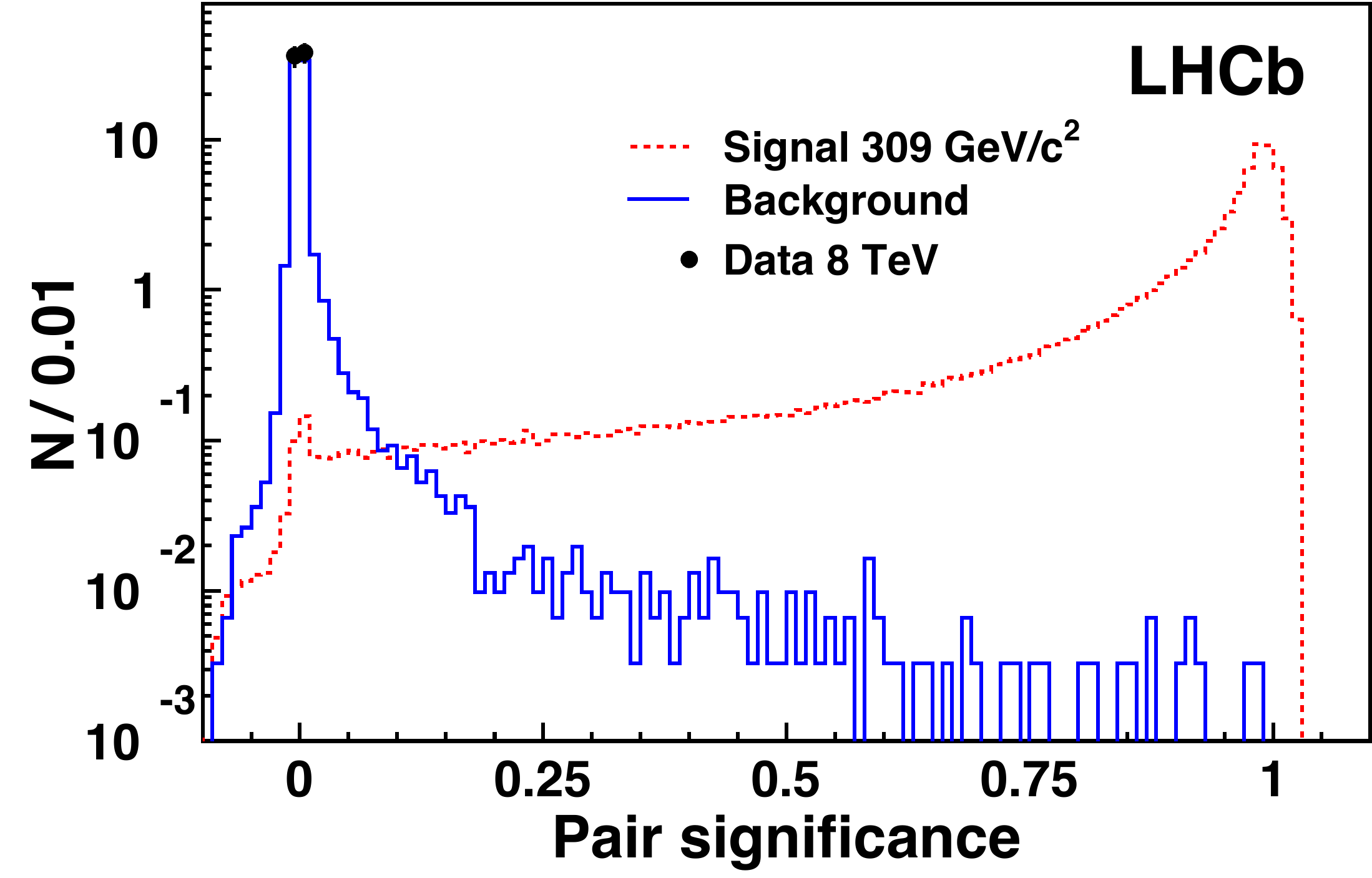}
\put(-116,80){\vector(0,-5){60}} 
  \caption{\small
Number of CMSP pairs, $N$, as a function of the pair significance.
The left and right figures correspond to the CMSP masses of 124 and 309\gevcc, respectively.
The black points with their
statistical uncertainty show the 7~TeV (top) and 8~TeV (bottom) data sets.
The red dashed histogram is
the expected shape from CMSP pairs and the blue histogram the background;
both are normalised to the number of events.  The arrows indicate the chosen selection criteria.
  }
\label{fig:significance-data} 
\end{center}
\end{figure}

Simulated events are used to train the ANN. The first three variables defined above are calibrated using
muons from $\Z / \gamma^{\star} \rightarrow \mu^+ \mu^-$ to ensure that simulation
agrees with data.  A total of 25k \Z events from the 2011 data set and 65k from the 2012 data set are used for the calibration.
In the \Z mass region the expected amount of signal is smaller than one event and cannot bias the procedure.
The DLLx variable is by far the most discriminating, and
its calibration procedure is presented in detail in Section~\ref{sec:riches}.

The ANN training is carried out independently for the 7 and 8\tev data sets, and for all the
CMSP mass hypotheses considered.
Figure~\ref{fig:4var_forANN} shows the distribution of the four ANN input variables for the 8\tev data set,
compared to the background and signal predictions; good agreement can be seen between data and
simulated background.

The discriminating variable is the product of the ANN outputs obtained for the two CMSP candidates. This ``pair
significance'' is shown in Fig.~\ref{fig:significance-data} for the 124 and 309\gevcc CMSP mass hypotheses
for both 7 and 8\tev data sets.
The requirement placed on the pair significance is determined by the value needed to achieve a 95\% signal
efficiency.
After the pair significance selection, the signal efficiency for candidate events in the \lhcb acceptance is
50\% for CMSPs with a mass of 124\gevcc, decreasing for increasing mass
to a minimum of 27\% at 309\gevcc.
The signal efficiency values for CMSPs in the acceptance,
after the ANN selection, are given in Table~\ref{tab:summary}.
After the full selection is applied, the dimuon background is suppressed by a factor of $10^{-5}$.


\section{CMSP identification with Cherenkov detectors}\label{sec:riches}
The present study uses the
Cherenkov radiation produced in the RICH detectors to identify CMSPs.
The Cherenkov momentum thresholds for muons, protons, and  CMSPs with masses of 124\gevcc and 309\gevcc,
are given in Table~\ref{tab:richpar} for the three radiators in the LHCb detectors.
Only CMSP candidates with momenta above 200\gevc are considered.
For this momentum range, 
particles with masses of the order of \mevcc to \gevcc,
have Cherenkov angles very close to the saturation value $\arccos (1/(n \beta))$, where $n$
is the refractive index of the medium.
The fraction of CMSPs with momentum above 2\tevc is negligible, and
the CMSPs are therefore expected not to produce Cherenkov radiation in the gaseous radiators.
Around half of the 124\gevcc CMSPs have a momentum above the Cherenkov threshold for aerogel,
and Cherenkov angles smaller than the saturation value.
This allows them to be separated from the background.
Only a few percent are expected to be in the momentum range $1.4-2.0\tevc$,
corresponding to Cherenkov angles from 0.225 to 0.234\rad.
It is possible to distinguish these angles from the saturation value of 0.242\rad in the aerogel as
the angular resolution is about 5.6\mrad.

As previously said, the variable DLLx has been introduced to identify  high momentum particles which do not radiate,
or have a Cherenkov angle which is too small to fit one of the five particle hypotheses,
$\pi$, $\mu$, $e$, K, p.
The DLLx value is positive for the momentum distributions of the CMSPs, for
all of the masses considered.\footnote{An anomalous signature in the RICH detectors could also be produced by tracks 
with wrongly assigned momenta.
This can happen if the particle has an absolute electric charge that is different from the proton charge.
For instance, a 1\gevcc proton-like particle is expected to produce Cherenkov light
in the two gaseous radiators when the measured momentum is above 30\gevc.
If the particle has one third of the proton charge, the measured momentum is overestimated by a factor of three and 
this will lead to an incorrect calculation of the Cherenkov emission.
}

\begin{table}[!tb]
\begin{center}
\small
\caption{\small Refractive indices and Cherenkov $\beta$ thresholds for the three radiators.
The momentum threshold is given for muons, protons, and 124 and 309\gevcc CMSPs.}
\label{tab:richpar}
  \begin{tabular}{c c c c c  c c }
\hline
&   & &  \multicolumn{4}{c}{$\ptot_{\rm thresh}$ (\gevc)} \\
\cline{4-7} \raisebox{1.5ex}{Radiator } & \raisebox{1.5ex}{$n$} & \raisebox{1.5ex}{$\beta_{\rm thresh}$} & $\mu$  &  p  & CMSP(124)  & CMSP(309)\\
\hline
Aerogel          & 1.03   & 0.9709  & 0.428   & 3.8     &502   & 1252\\
$\rm C_4F_{10}$   & 1.0014 & 0.9985  & 2.00    & 17.7    &2342   & 5069\\
$\rm CF_4$       & 1.0005 & 0.9995  & 3.34    & 29.7    &3921  & 9767\\
\hline
\end{tabular}
\end{center}
\end{table}

Simulated dimuon background events and CMSP signal samples used to train the ANN
are first validated with data.

\begin{figure}[!tb]
\begin{center}
\includegraphics[width=0.49\linewidth]{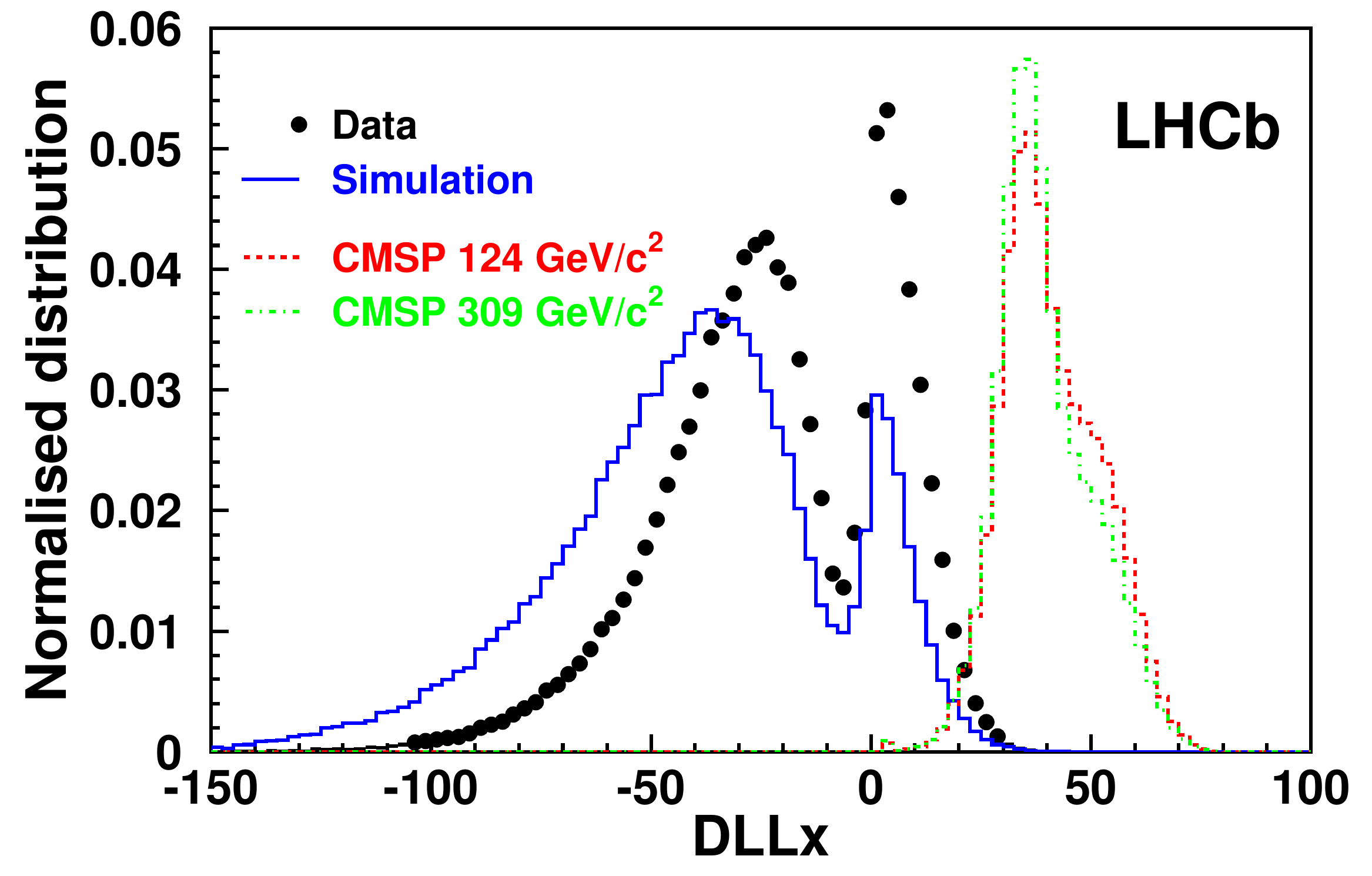}  \put(-30,80){(a)} 
\includegraphics[width=0.49\linewidth]{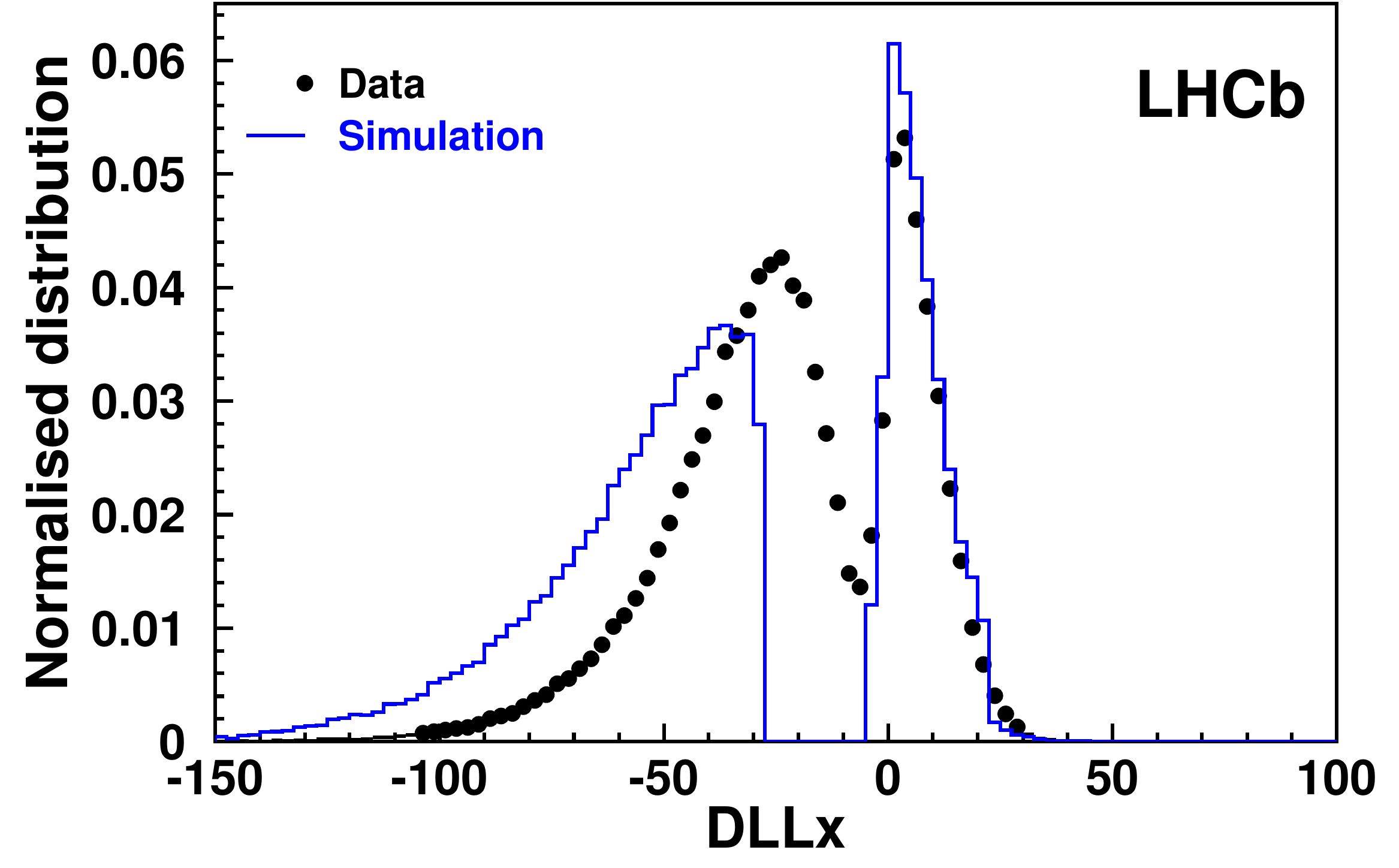}  \put(-30,80){(b)}
\caption{\small (a) DLLx distributions from muons selected from $Z$ decays in 7~TeV data (black points) and simulation (blue histogram).
  The expected shape from 124\gevcc and 309\gevcc CMSPs are also shown.
  (b) The simulated DLLx values are shifted in such a way that data and simulation have
  the same fraction of entries with $\mathrm{DLLx}>-5$.}
\label{fig:DLLbt-cali}
\end{center}
\end{figure}

The study of the background samples is performed on a set of muons above the Cherenkov threshold
and selected from \Z decays.
Such events have an event topology and kinematics that are very close to those of the dimuon background
expected in the CMSP analysis.
The DLLx distribution is shown in Fig.~\ref{fig:DLLbt-cali}~(a) for muons from data and simulated \Z decays.
For illustration, the expected signal shapes for CMSPs with masses of 124 and 309\gevcc are superimposed.
The small difference is due to a change in the underlying event and
some light from the aerogel for the 124\gevcc case.
A clear separation between the signal and background muon DLLx distributions is observed.
The difference in the data and simulated muon distributions is mainly due to the lack of precision in the mapping
of the photon detection efficiency in the RICH system.
In particular, the peak at $\mathrm{DLLx}>-5$ is produced by
the decrease of the photon detection efficiency when approaching boundaries in the RICH modules.
The simulation only partially reproduces this behaviour and the number of candidates above $\mathrm{DLLx}=-5$ is
too low by around a factor of two.
%
%
To compensate for this, 15\% of simulated muon events with DLLx falling
close to zero have been shifted by an {\em ad hoc} value to obtain the best agreement between data
and simulation in the $\mathrm{DLLx}>-5$ region,
resulting in the distribution shown in Fig.~\ref{fig:DLLbt-cali}~(b).
It is expected that a correct efficiency map should produce such a shift, moving above zero the
slightly negative DLLx values.
This set of simulated background events is used to train the ANN.
Note that only candidates with $\mathrm{DLLx}>-5$ are used in the ANN.
In order to assess the systematic uncertainty associated to this correction method,
two other procedures are considered. In the first procedure, the ANN training is performed
on the original background simulated data set.
In the second,  the DLLx values for each muon are randomly chosen following a set of
templates inferred from the DLLx data distributions as function of \ptot and $\eta$.
Despite the fact that this operation is done in bins of \ptot and $\eta$,
it is obvious that most of the correlation is lost in the randomisation process.
The three methods are found to provide the same final discrimination power
and their contributions to the systematic uncertainties are small.
This is due to the strong separation between signal and background that is provided intrinsically by the DLLx variable.

The validation of the signal sample is more complex due to the absence of a SM process that can be used for calibration.
The quality of the simulation was studied using protons from  $\Lz \rightarrow {\rm p} \pi$ decays
with a velocity below the Cherenkov threshold.
The differences between the data and simulation for these protons are extrapolated to the CMSP
kinematics using a fast simulation method,
and the contribution to the systematic uncertainty estimated.

The proton is below the Cherenkov threshold in all of the RICH radiators for $\ptot<3.8\gevc$, and above the threshold
for $\ptot \gtrsim 30\gevc$.
Pairs of opposite-charge tracks are selected from data and from simulated events passing a minimum bias trigger.
The pair must combine to form a particle with a mass compatible with the known mass of the \Lz baryon,
and the reconstructed vertex must be more than 3\mm from the beam axis.

\begin{figure}[!tb]
\begin{center}
  \includegraphics[width=0.5\linewidth]{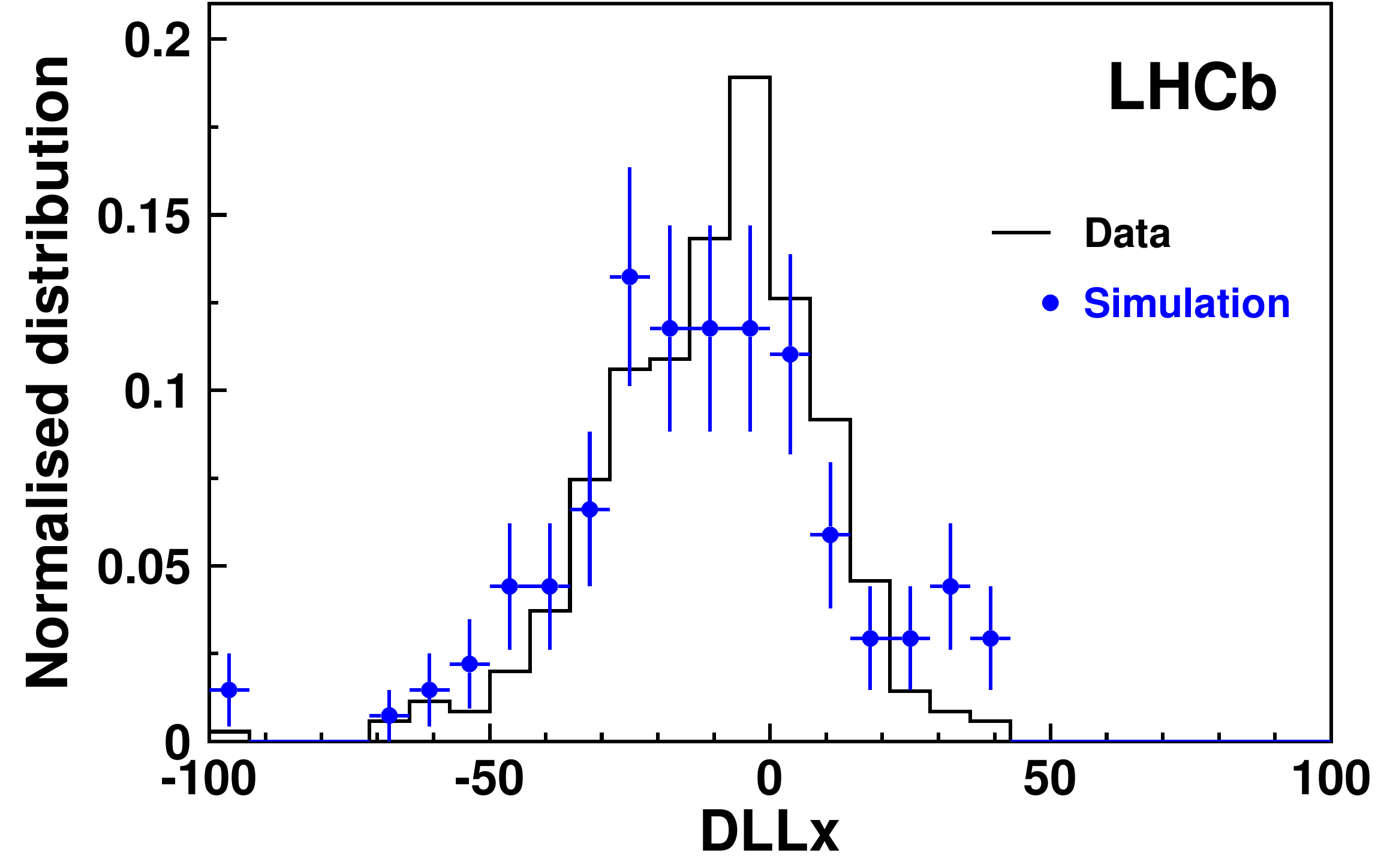} \put(-160,120){(a)}\\
  \includegraphics[width=0.5\linewidth]{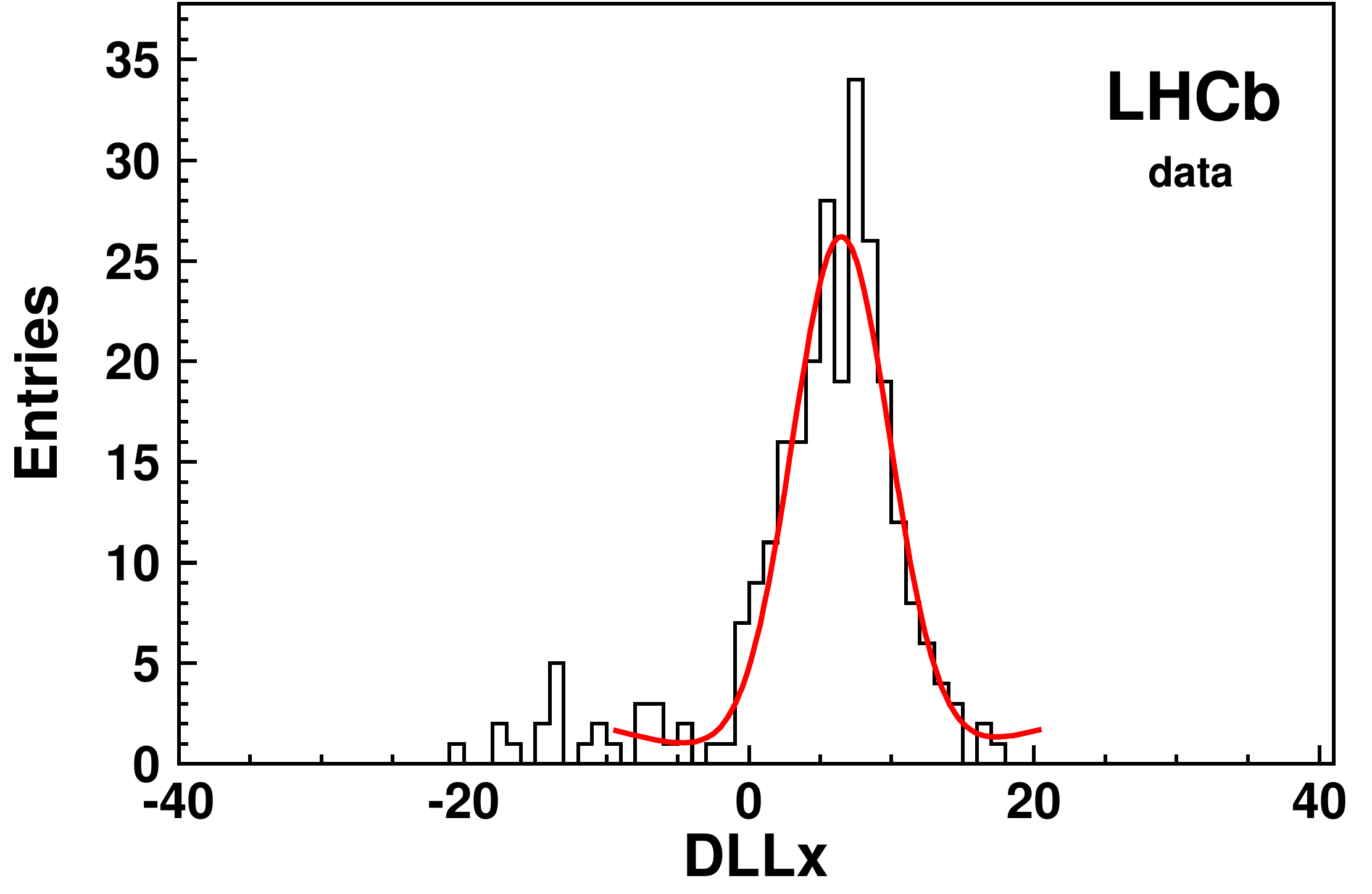} \put(-160,120){(b)}
  \includegraphics[width=0.5\linewidth]{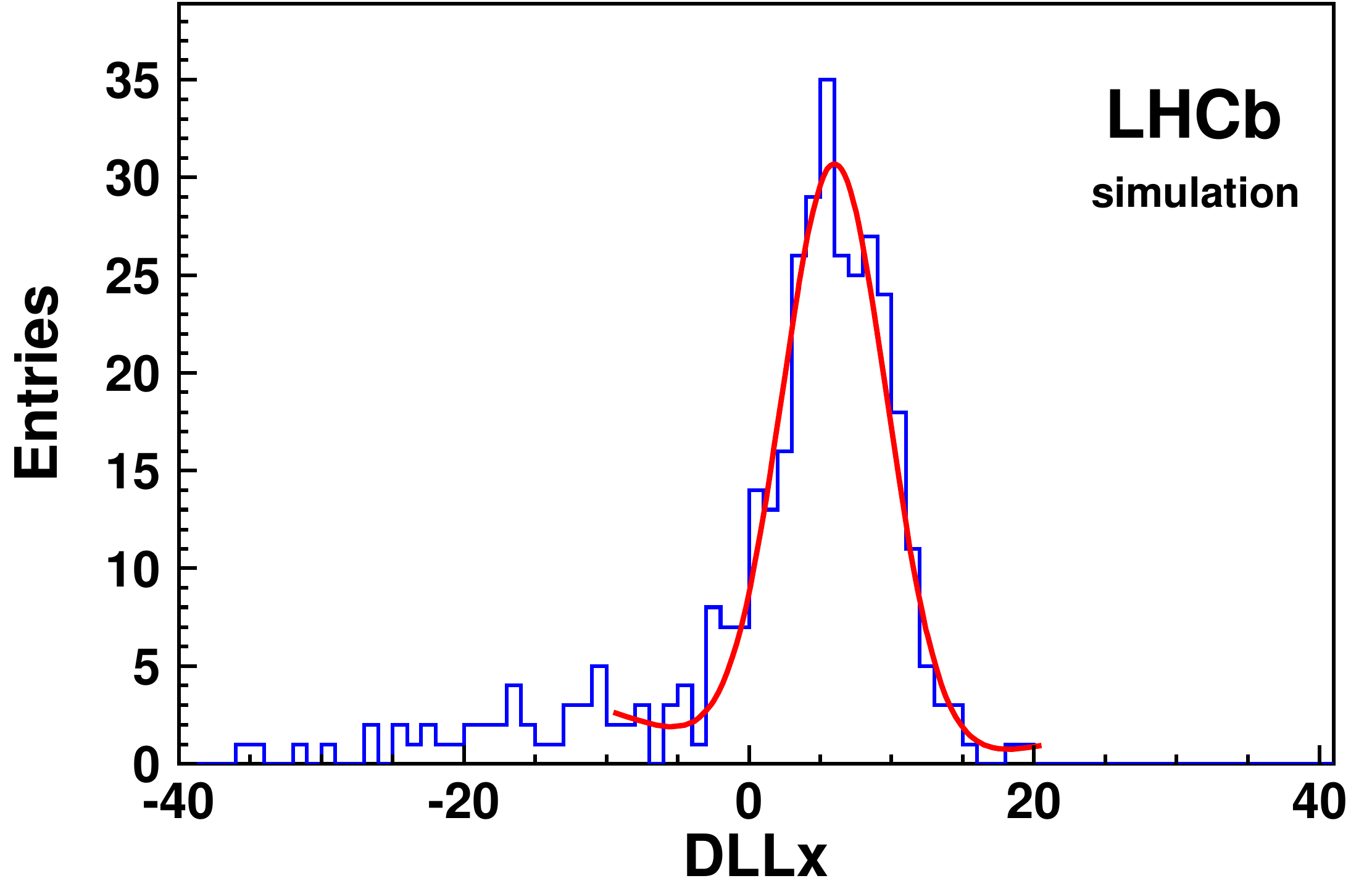} \put(-160,120){(c)}
    \caption{\small
    The DLLx variable for: (a) protons selected above the Cherenkov threshold, $\ptot > 30$\gevc, in 7~TeV data (black line) and simulation
    (blue points); (b) and (c) below-threshold protons, $\ptot < 3.8\gevc$, from data and simulation, respectively.
    The results from the fit of a Gaussian function plus polynomial to the below-threshold proton distributions are shown by the red curves.
}
\label{fig:lambda-2}
\end{center}
\end{figure}

Samples of protons below and above Cherenkov threshold are obtained by choosing the
momentum regions below 3.8\gevc and above 30\gevc, respectively.
Figure~\ref{fig:lambda-2} shows the corresponding DLLx distributions,
indicating a reasonable agreement between
the DLLx distributions from data collected at 7\tev and simulation.
A Gaussian function plus a polynomial, to account for the tail at low DLLx, is fitted to 
the DLLx distributions for below-Cherenkov-threshold particles.
The mean and width of the Gaussian functions are $6.5\pm0.3$ and $3.4\pm0.3$ for data,
and $6.0\pm0.3$ and $3.6\pm0.4$ for simulation.
The DLLx value in data is $0.5\pm0.4$ units higher, which may indicate that there is a
lower photon detection efficiency in data compared to simulation.  
A maximal deviation of $\pm 1$ DLLx units is considered in the following to assess the systematic
effects. The extrapolation from the low momentum proton result to the CMSP regime is
made using a fast simulation.

In addition to the \geant-based full simulation, a fast simulation describing the main
features of the RICH measurement process is also used.
This allows the impact of varying
parameters and the algorithms to be studied in a more efficient way.
The fast simulation generates a target particle (a proton from \Lz decays, a muon or a CMSP)  with
a momentum distribution representing the phenomenon under study.
The underlying event is represented by a number of pions with a momentum distribution obtained from
minimum bias events.
The simulation of Cherenkov emission in the radiators is then performed for each particle.
The number of Cherenkov photons
generated by a particle of velocity $\beta$ follows a Poisson distribution of average
${\rm N_0} (\beta^2 n^2 -1)/(\beta^2 (n^2 -1))$,
where $\rm N_0$ is the maximal number of photons for a saturated ring and $n$ is the refractive index.
The ring has an average radius corresponding to the expected Cherenkov angle and
a Gaussian profile of width $\sigma_c$ representing the angular resolution of the detector.
Finally, random noise is added using the probability for a pixel to fire, $\rm prob_{noise}$.
The nominal values of the parameters used in the fast simulation are given in Table~\ref{tab:rich-sim}.
The event log-likelihood for each target particle hypothesis is
\begin{equation}
\rm LL = -\sum_{\text{pixel}\, i}^{\text{all pixels}}\nu_i + \sum_{\text{pixel}\, i}^{\text{active pixels}} \ln{(e^{\nu_i}-1)}
\end{equation}
where $\nu_i$ is the probability to have
photons in the pixel $i$,
including the random background. Note that the formula is valid for
the binary readout implemented in the RICH electronics.
The centre of each Cherenkov ring is defined by the true particle direction.
The DLL values are subsequently computed.

\begin{table}[!tb]
\begin{center}
\caption{\small Nominal values of the parameters used in the fast simulation.}
\begin{tabular}{c c c c}
  \hline
  parameter             & aerogel & $\rm C_4F_{10}$ & $\rm CF_{4}$ \\
  \hline
  $n$                   & 1.03    & 1.0014     & 1.0005   \\
  $\rm N_0$             & 8      &  28        & 24       \\
  $\sigma_c$ [mrad]     & 5.6     &  1.6       & 0.7      \\
  $\rm prob_{noise}$     & 3\%     &  3\%       & 3\%      \\
  \hline
\end{tabular}
\label{tab:rich-sim}
\end{center}
\end{table}

\begin{figure}[!tb]
\begin{center}
  \includegraphics[width=0.48\linewidth]{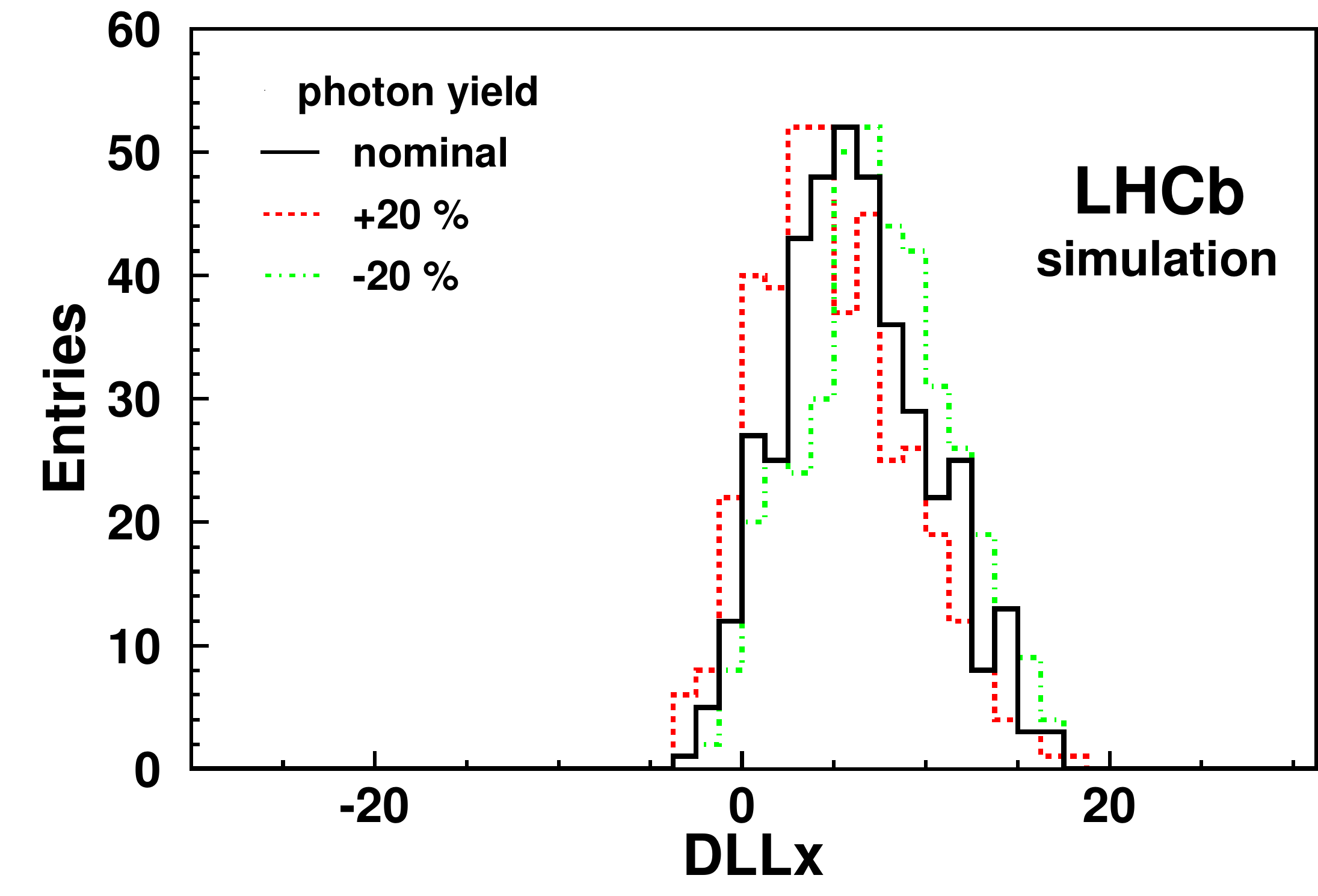} \put(-30,85){(a)}
\hspace{1mm}  \includegraphics[width=0.48\linewidth]{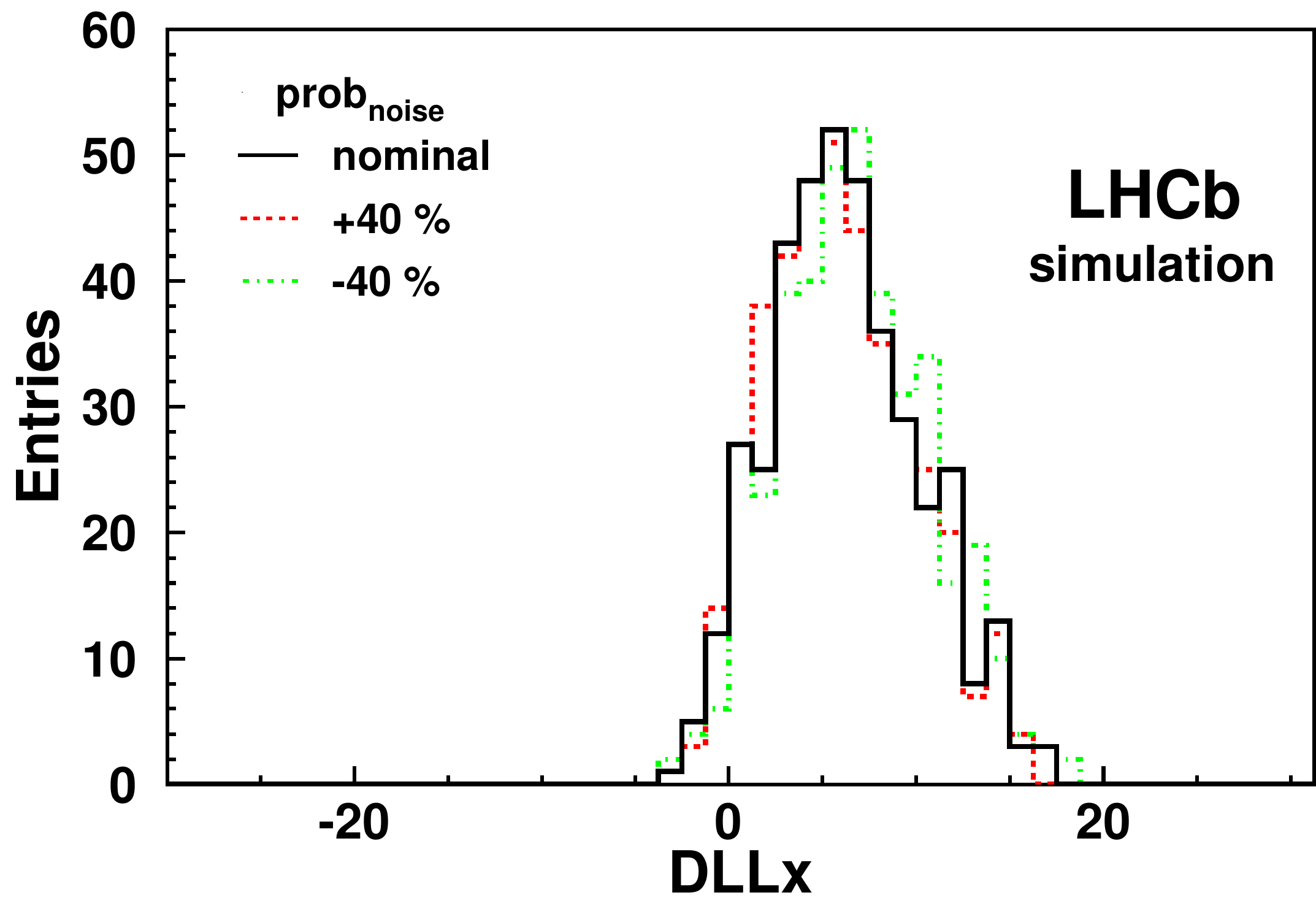}    \put(-30,85){(b)}
  \caption{\small DLLx for protons with $\ptot < 3.8\gevc$.
    In black the result with nominal simulation parameters are shown.
    (a) The red, dashed, and green, dash-dot, plots are for a change by +20\% and -20\% of
    the photon yield;
    (b) is for a change by +40\% (red, dashed) and -40\% (green, dash-dot) of the random noise probability. 
}
\label{fig:rich-sim-dllbt-proton}
\end{center}
\end{figure}

Simulated distributions for protons from \Lz decays with $\ptot < 3.8\gevc$ are shown in
Fig.~\ref{fig:rich-sim-dllbt-proton}.~The average DLLx is 6.2, with an RMS of 4.0,
for a fast simulation made using the nominal parameters.
Figure~\ref{fig:rich-sim-dllbt-proton}~(a) also shows the distributions after
varying the detection efficiency by $\pm 20$\%.
The corresponding distributions are shifted by $\mp 1.2$ units.
The random noise probability was changed by $\pm 40$\% from its nominal value which
produces negligible variation as seen in Fig.~\ref{fig:rich-sim-dllbt-proton}~(b). 
This study shows that a variation of  $\pm 1$ DLLx units is 
obtained by changing the photon detection efficiency by  $\mp 15$\%.
A variation of the same size can be obtained by changing the
angular resolution $\sigma_c$ by 50\%.

The DLLx distributions for CMSPs obtained from 
the fast simulation with nominal parameters
are consistent with those obtained from full simulation.
Changing the photon detection efficiency by $\pm 15$\%, as inferred from the \Lz study,
a variation of  $\pm 2$ DLLx units is obtained.
An identical result can be obtained by changing the angular resolution.

In summary, an uncertainty of two DLLx units is inferred
from the comparison of below-threshold protons in data and simulation when using \Lz decays.
The extrapolation to the CMSP regime is obtained by varying the simulation
parameters and leads to an uncertainty of four DLLx units on the average DLLx value for such particles.

\section{Uncertainties and results} \label{results}

After the ANN selection the signal prediction for the chosen model is 2.5, 0.9 and 0.3 events for the 124, 154 and 185\gevcc
\stau masses, and below 0.1 for the other mass values.
The expected background is negligible, less than 0.02 events for all the mass hypotheses.  \par

A summary of the systematic uncertainties is given in Table~\ref{tab:sys}.
The total systematic uncertainties are approximately 5\% for the signal yield
and 50\% for the background yield.

\begin{table}[tb]
 \begin{center}
   \caption{\small Systematic uncertainties (in \%) for the selection of the signal of CMSP pairs
     and on the background retention.
     Where relevant, the lower value corresponds to the CMSP mass of 124\gevcc and the higher one to 309\gevcc.}
\begin{tabular}[h]{ l c c c c }
\hline
& \multicolumn{2}{c }{7\tev} & \multicolumn{2}{c }{8\tev}\\
\cline{2-5}                & Signal         & Background      & Signal       & Background\\
                           & efficiency     & retention       & efficiency    & retention \\
\hline

Luminosity                &      1.7     &      1.7      &      1.2     &     1.2    \\
Trigger, reconstruction   &      2.0     &      2.0      &      2.0     &     2.0       \\
Delayed signals           &   2.2--3.7   &       0       &    2.2--3.7  &     0       \\
IP calibration            &       0      &      0.7      &        0     &     0.7     \\
Hadron electron rejection &       0      &      15       &        0     &     15      \\
Neural Network            &      2.9     &      50.0     &      2.9     &     50.0     \\
\hline
Total syst. uncertainty   &     4.5--5.4 &      52.3     &    4.3--5.2   &     52.3 \\
\hline
 \end{tabular}
\label{tab:sys}
\end{center}
\end{table}

Two methods are used to determine the luminosity: a Van der Meer
 scan and a beam-gas imaging method~\cite{LHCb-PAPER-2014-047}.
 The  uncertainties on the integrated luminosities are 1.7\%  for the 7\tev data set and 1.2\%  for the 8\tev data set.

The efficiency for triggering, reconstructing and identifying high-\pt muons has been studied in detail for the
\lhcb $\Z$ and $\W$ boson cross-section
measurements~\cite{LHCb-PAPER-2015-001,LHCb-PAPER-2014-033},
and the agreement between data and simulation was found to be better than 2\%.
This percentage is taken as the corresponding uncertainty for this analysis.

A further efficiency uncertainty is considered due to the delayed signals in the tracking and muon
systems. The timing precision affects
  the amplitude recorded by the front-end electronic boards and the measurement of the drift time in the straw tubes.
  The effect on the efficiency due to a timing uncertainty of $\pm 1$~ns
  is determined from simulation as a function of the $\beta$ of the particle.
  Subsequently, a weighted average of the uncertainty is obtained from the $\beta$ distributions for each mass hypothesis,
  providing values varying from 2.2\% to 3.7\% for CMSP masses from 124\gevcc to 309\gevcc.

  The comparison of the IP distributions in data and simulated $\Z/\gamma^{\star} \rightarrow \mu^+ \mu^-$ events
  indicates a maximal discrepancy of  $\pm 5~\mu$m. By changing the requirement
  on the IP parameter by this amount, the
  corresponding  efficiency variation is $\pm 0.7$\% for the background and negligible for the signal.

  The hadron and electron rejection is affected by the calibration of the calorimeters.
  From the comparison of 
  $\Z/\gamma^{\star} \rightarrow \mu^+ \mu^-$ decays in data and simulation
  a relative uncertainty of 10\% is inferred.
  This translates into a 15\% change on the background yield, while the signal is almost unaffected.

  The training of the ANN is affected by the uncertainty on the background and signal models.
  A 2.7\% contribution to the signal efficiency uncertainty
  is associated with the calibration procedures, determined by the comparison
  of data and simulation for
  $\Z/\gamma^{\star} \rightarrow \mu^+ \mu^-$ and  $\Lz \to \mathrm{p} \pi$  decays.
  Error propagation is performed by modifying the ANN training sets, while keeping the test
  sets and the pair significance selection fixed.
  Adding the uncertainties in quadrature with the statistical uncertainty of 1\%, gives
  a total of 2.9\%.

  The ANN selection leaves a very small amount of simulated background.
  The binomial uncertainty on the background retention is large, at approximately 50\%.
  This value is assigned as the uncertainty on the background selection efficiency.

  As already stated, the acceptance A is affected by model uncertainties in the range
  from 5\% to 9\% for \stau  mass from 124\gevcc to 309\gevcc. 
  In addition, the choice of the PDF affects the efficiency by modifying the momentum
  of the CMSP. By scanning various PDFs, we have found that this effect is small,
  not larger that 0.4\%, for all the models.

\begin{figure}[h!]
\begin{center}
\hspace{-10mm}\includegraphics[width=0.65\linewidth]{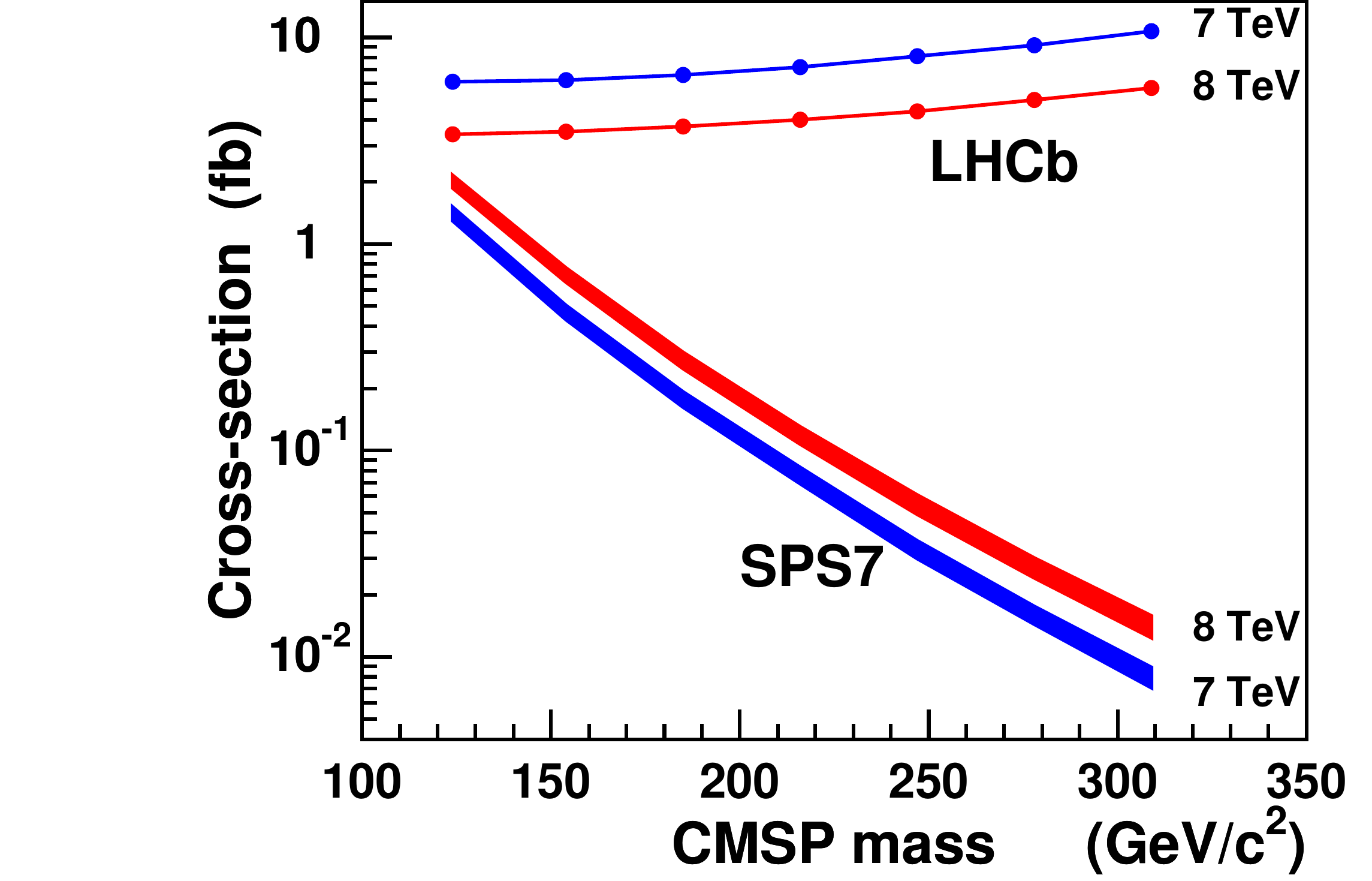}
\caption{\small
  Upper limits at 95\% CL on the cross-sections for the pair production of CMSPs in the LHCb acceptance (points)
  and the corresponding predictions assuming the Drell-Yan production of \stau
  (bands representing $\pm 1 \sigma$ uncertainty) with SPS7 parameters, 
  for proton-proton collisions as a function of the CMSP mass
  at \sqs=7 and 8~TeV.
}
\label{fig:limit_inLHCb}
\end{center}
\end{figure}

The cross-section upper limits are computed using the Feldman-Cousins method~\cite{F-C}
for zero observed candidates,
taking into account the expected number of background events and the uncertainties~\cite{pole}.
The predicted amount of background is so small that it has no sizeable effect on the result.
The upper limits at a 95\% confidence level (CL) for CMSP pair production in the LHCb geometrical
acceptance at \sqs = 7 and 8\tev
are listed in Table~\ref{tab:limits} and shown in Fig.~\ref{fig:limit_inLHCb} 
together with the theoretical cross-sections calculated for the particular
model described in Section~\ref{sec:gen-stau}.  \par

\begin{table}[tb]
\begin{center}
\caption{\small
  Cross-section upper limits at 95\% CL  for CMSP pair production in the LHCb acceptance
  in the 7 and 8~TeV. }
\begin{tabular}{c c c  }
\hline
$m_{\rm CMSP}$  & \multicolumn{2}{c}{\small Upper limit (fb)}   \\
   \cline{2-3} (\gevcc) &  7\tev & 8\tev \\
\hline
124    &        6.1 &        3.4  \\
154    &        6.2 &        3.5  \\
185    &        6.6 &        3.7  \\
216    &        7.2 &        4.0  \\ 
247    &        8.1 &        4.4  \\ 
278    &        9.2 &        5.0  \\
309    &       10.7 &        5.7  \\
\hline
\end{tabular}
\label{tab:limits}
\end{center}
\end{table}

\section{Conclusions}

A search for pairs of long-lived charged particles,
with masses in the range 124--309\gevcc,
using \stau pairs predicted by the mGMSB model as a benchmark scenario,
is performed using data from proton-proton collisions at 7 and 8\tev,
corresponding to an integrated luminosity of 3.0 $\invfb$,
collected with the LHCb detector in
the forward pseudorapidity range $1.8 <\eta <4.9$.
The candidates are assumed to interact only through the electroweak interaction in the detector.
Hence, they behave like heavy muons and their main signature is the absence of a signal in the RICH detectors.
The detection efficiency is limited to particles with $\beta > 0.8$ due to the acceptance in time after beam
crossing.
The main background contribution
comes from  $\Z/\gamma^{\star} \to \mu^+\mu^-$ and is reduced to less than $\sim 0.02$ events.
No events have been observed.
Upper limits are set on the Drell-Yan CMSP pair production cross-section.
For proton-proton collisions at 
\sqs= 7\tev, the 95\% CL upper limits for the production cross-section of a pair of CMSPs
in the LHCb acceptance vary from 6.1~fb for a mass of 124\gevcc up to 10.7~fb for a mass of 309\gevcc.
At \sqs= 8\tev, they vary from 3.4~fb to 5.7~fb for the same masses.

In \lhcb the identification of CMSPs relies on the
energy deposited in the subdetectors,
the main discrimination power being provided by the RICH system.
Together with the forward pseudorapidity coverage, this unique
feature allows \lhcb to complement the searches undertaken by the
central detectors at the Tevatron and LHC.

\section*{Acknowledgements}
 
\noindent We express our gratitude to our colleagues in the CERN
accelerator departments for the excellent performance of the LHC. We
thank the technical and administrative staff at the LHCb
institutes. We acknowledge support from CERN and from the national
agencies: CAPES, CNPq, FAPERJ and FINEP (Brazil); NSFC (China);
CNRS/IN2P3 (France); BMBF, DFG, HGF and MPG (Germany); INFN (Italy); 
FOM and NWO (The Netherlands); MNiSW and NCN (Poland); MEN/IFA (Romania); 
MinES and FANO (Russia); MinECo (Spain); SNSF and SER (Switzerland); 
NASU (Ukraine); STFC (United Kingdom); NSF (USA).
The Tier1 computing centres are supported by IN2P3 (France), KIT and BMBF 
(Germany), INFN (Italy), NWO and SURF (The Netherlands), PIC (Spain), GridPP 
(United Kingdom).
We are indebted to the communities behind the multiple open 
source software packages on which we depend. We are also thankful for the 
computing resources and the access to software R\&D tools provided by Yandex LLC (Russia).
Individual groups or members have received support from 
EPLANET, Marie Sk\l{}odowska-Curie Actions and ERC (European Union), 
Conseil g\'{e}n\'{e}ral de Haute-Savoie, Labex ENIGMASS and OCEVU, 
R\'{e}gion Auvergne (France), RFBR (Russia), XuntaGal and GENCAT (Spain), Royal Society and Royal
Commission for the Exhibition of 1851 (United Kingdom).

\addcontentsline{toc}{section}{References}
\setboolean{inbibliography}{true}
\bibliographystyle{LHCb}
\bibliography{main,LHCb-PAPER,LHCb-CONF,LHCb-DP}

\ifx\mcitethebibliography\mciteundefinedmacro
\PackageError{LHCb.bst}{mciteplus.sty has not been loaded}
{This bibstyle requires the use of the mciteplus package.}\fi
\providecommand{\href}[2]{#2}
\begin{mcitethebibliography}{10}
\mciteSetBstSublistMode{n}
\mciteSetBstMaxWidthForm{subitem}{\alph{mcitesubitemcount})}
\mciteSetBstSublistLabelBeginEnd{\mcitemaxwidthsubitemform\space}
{\relax}{\relax}

\bibitem{Dimopoulos}
S.~Dimopoulos, S.~D. Thomas, and J.~D. Wells,
  \ifthenelse{\boolean{articletitles}}{\emph{{Sparticle spectroscopy and
  electroweak symmetry breaking with gauge mediated supersymmetry breaking}},
  }{}\href{http://dx.doi.org/10.1016/S0550-3213(97)00030-8}{Nucl.\ Phys.\
  \textbf{B488} (1997) 39}, \href{http://arxiv.org/abs/hep-ph/9609434}{{\tt
  arXiv:hep-ph/9609434}}\relax
\mciteBstWouldAddEndPuncttrue
\mciteSetBstMidEndSepPunct{\mcitedefaultmidpunct}
{\mcitedefaultendpunct}{\mcitedefaultseppunct}\relax
\EndOfBibitem
\bibitem{gmsb_giudice}
G.~F. Giudice and R.~Rattazzi,
  \ifthenelse{\boolean{articletitles}}{\emph{{Theories with gauge-mediated
  supersymmetry breaking}},
  }{}\href{http://dx.doi.org/10.1016/S0370-1573(99)00042-3}{{Phys.\ Rep.\ }
  \textbf{332} (2011) 419}, \href{http://arxiv.org/abs/hep-ph/9801271}{{\tt
  arXiv:hep-ph/9801271}}\relax
\mciteBstWouldAddEndPuncttrue
\mciteSetBstMidEndSepPunct{\mcitedefaultmidpunct}
{\mcitedefaultendpunct}{\mcitedefaultseppunct}\relax
\EndOfBibitem
\bibitem{susy_martin}
S.~P. Martin, \ifthenelse{\boolean{articletitles}}{\emph{{A supersymmetry
  primer}}, }{}\href{http://arxiv.org/abs/hep-ph/9709356}{{\tt
  arXiv:hep-ph/9709356}}\relax
\mciteBstWouldAddEndPuncttrue
\mciteSetBstMidEndSepPunct{\mcitedefaultmidpunct}
{\mcitedefaultendpunct}{\mcitedefaultseppunct}\relax
\EndOfBibitem
\bibitem{ALEPH}
ALEPH collaboration, V.~Barate {\em et~al.},
  \ifthenelse{\boolean{articletitles}}{\emph{Search for pair-production of
  long-lived heavy charged particles in $e^+e^-$ annihilation},
  }{}\href{http://dx.doi.org/10.1016/S0370-2693(97)00715-6}{Phys.\ Lett.\
  \textbf{B405} (1997) 379}, \href{http://arxiv.org/abs/hep-ex/9706013}{{\tt
  arXiv:hep-ex/9706013}}\relax
\mciteBstWouldAddEndPuncttrue
\mciteSetBstMidEndSepPunct{\mcitedefaultmidpunct}
{\mcitedefaultendpunct}{\mcitedefaultseppunct}\relax
\EndOfBibitem
\bibitem{DELPHI}
DELPHI collaboration, P.~Abreu {\em et~al.},
  \ifthenelse{\boolean{articletitles}}{\emph{Search for heavy stable and
  long-lived particles in $e^+e^-$ collisions at \sqs=189 \gev},
  }{}\href{http://dx.doi.org/10.1016/S0370-2693(00)00265-3}{Phys.\ Lett.\
  \textbf{B478} (2000) 65}, \href{http://arxiv.org/abs/hep-ex/0103038}{{\tt
  arXiv:hep-ex/0103038}}\relax
\mciteBstWouldAddEndPuncttrue
\mciteSetBstMidEndSepPunct{\mcitedefaultmidpunct}
{\mcitedefaultendpunct}{\mcitedefaultseppunct}\relax
\EndOfBibitem
\bibitem{L3}
L3 collaboration, P.~Achard {\em et~al.},
  \ifthenelse{\boolean{articletitles}}{\emph{Search for heavy neutral and
  charged leptons in $e^+e^-$ annihilation at \lep},
  }{}\href{http://dx.doi.org/10.1016/S0370-2693(01)01005-X}{Phys.\ Lett.\
  \textbf{B517} (2001) 75}, \href{http://arxiv.org/abs/hep-ex/0107015}{{\tt
  arXiv:hep-ex/0107015}}\relax
\mciteBstWouldAddEndPuncttrue
\mciteSetBstMidEndSepPunct{\mcitedefaultmidpunct}
{\mcitedefaultendpunct}{\mcitedefaultseppunct}\relax
\EndOfBibitem
\bibitem{OPAL}
OPAL collaboration, G.~Abbiendi {\em et~al.},
  \ifthenelse{\boolean{articletitles}}{\emph{Search for stable and long-lived
  massive charged particles in $e^+e^-$ collisions at \sqs=130--209 \gev},
  }{}\href{http://dx.doi.org/10.1016/S0370-2693(03)00639-7}{Phys.\ Lett.\
  \textbf{B572} (2003) 8}, \href{http://arxiv.org/abs/hep-ex/0305031}{{\tt
  arXiv:hep-ex/0305031}}\relax
\mciteBstWouldAddEndPuncttrue
\mciteSetBstMidEndSepPunct{\mcitedefaultmidpunct}
{\mcitedefaultendpunct}{\mcitedefaultseppunct}\relax
\EndOfBibitem
\bibitem{H1}
H1 collaboration, G.~Aktas {\em et~al.},
  \ifthenelse{\boolean{articletitles}}{\emph{Measurement of anti-deuteron
  photoproduction and a search for heavy stable charged particles at
  \texttt{HERA}}, }{}\href{http://dx.doi.org/10.1140/epjc/s2004-01978-x}{Eur.\
  Phys.\ J.\  \textbf{C36} (2004) 413},
  \href{http://arxiv.org/abs/hep-ex/0403056}{{\tt arXiv:hep-ex/0403056}}\relax
\mciteBstWouldAddEndPuncttrue
\mciteSetBstMidEndSepPunct{\mcitedefaultmidpunct}
{\mcitedefaultendpunct}{\mcitedefaultseppunct}\relax
\EndOfBibitem
\bibitem{limit_D0}
D0 collaboration, V.~M. Abazov {\em et~al.},
  \ifthenelse{\boolean{articletitles}}{\emph{{Search for long-lived charged
  massive particles with the D0 detector}},
  }{}\href{http://dx.doi.org/10.1103/PhysRevLett.102.161802}{{Phys.\ Rev.\
  Lett.\ } \textbf{102} (2009) 161802},
  \href{http://arxiv.org/abs/0809.4472}{{\tt arXiv:0809.4472}}\relax
\mciteBstWouldAddEndPuncttrue
\mciteSetBstMidEndSepPunct{\mcitedefaultmidpunct}
{\mcitedefaultendpunct}{\mcitedefaultseppunct}\relax
\EndOfBibitem
\bibitem{limit_CDF}
CDF collaboration, T.~Aaltonen {\em et~al.},
  \ifthenelse{\boolean{articletitles}}{\emph{{Search for long-lived massive
  charged particles in 1.96 TeV $p\overline{p}$ collisions}},
  }{}\href{http://dx.doi.org/10.1103/PhysRevLett.103.021802}{{Phys.\ Rev.\
  Lett.\ } \textbf{103} (2009) 021802},
  \href{http://arxiv.org/abs/0902.1266}{{\tt arXiv:0902.1266}}\relax
\mciteBstWouldAddEndPuncttrue
\mciteSetBstMidEndSepPunct{\mcitedefaultmidpunct}
{\mcitedefaultendpunct}{\mcitedefaultseppunct}\relax
\EndOfBibitem
\bibitem{limit_atlas_n}
ATLAS collaboration, G.~Aad {\em et~al.},
  \ifthenelse{\boolean{articletitles}}{\emph{{Searches for heavy long-lived
  charged particles with the ATLAS detector in proton-proton collisions at
  $\sqrt{s} = 8$ TeV}},
  }{}\href{http://dx.doi.org/10.1007/JHEP01(2015)068}{{JHEP} \textbf{01} (2015)
  68}, \href{http://arxiv.org/abs/{1411.​6795}}{{\tt
  arXiv:{1411.​6795}}}\relax
\mciteBstWouldAddEndPuncttrue
\mciteSetBstMidEndSepPunct{\mcitedefaultmidpunct}
{\mcitedefaultendpunct}{\mcitedefaultseppunct}\relax
\EndOfBibitem
\bibitem{limit_cms}
CMS collaboration, S.~Chatrchyan {\em et~al.},
  \ifthenelse{\boolean{articletitles}}{\emph{{Searches for long-lived particles
  in $pp$ collisions at $\sqrt{s} = 7$ and 8 TeV}},
  }{}\href{http://dx.doi.org/10.1007/JHEP07(2013)122}{{JHEP} \textbf{07} (2013)
  122}, \href{http://arxiv.org/abs/1305.0491}{{\tt arXiv:1305.0491}}\relax
\mciteBstWouldAddEndPuncttrue
\mciteSetBstMidEndSepPunct{\mcitedefaultmidpunct}
{\mcitedefaultendpunct}{\mcitedefaultseppunct}\relax
\EndOfBibitem
\bibitem{Alves:2008zz}
LHCb collaboration, A.~A. Alves~Jr.\ {\em et~al.},
  \ifthenelse{\boolean{articletitles}}{\emph{{The \lhcb detector at the LHC}},
  }{}\href{http://dx.doi.org/10.1088/1748-0221/3/08/S08005}{JINST \textbf{3}
  (2008) S08005}\relax
\mciteBstWouldAddEndPuncttrue
\mciteSetBstMidEndSepPunct{\mcitedefaultmidpunct}
{\mcitedefaultendpunct}{\mcitedefaultseppunct}\relax
\EndOfBibitem
\bibitem{LHCb-DP-2014-002}
LHCb collaboration, R.~Aaij {\em et~al.},
  \ifthenelse{\boolean{articletitles}}{\emph{{LHCb detector performance}},
  }{}\href{http://dx.doi.org/10.1142/S0217751X15300227}{Int.\ J.\ Mod.\ Phys.\
  \textbf{A30} (2015) 1530022}, \href{http://arxiv.org/abs/1412.6352}{{\tt
  arXiv:1412.6352}}\relax
\mciteBstWouldAddEndPuncttrue
\mciteSetBstMidEndSepPunct{\mcitedefaultmidpunct}
{\mcitedefaultendpunct}{\mcitedefaultseppunct}\relax
\EndOfBibitem
\bibitem{LHCb-DP-2014-001}
R.~Aaij {\em et~al.}, \ifthenelse{\boolean{articletitles}}{\emph{{Performance
  of the LHCb Vertex Locator}},
  }{}\href{http://dx.doi.org/10.1088/1748-0221/9/09/P09007}{JINST \textbf{9}
  (2014) P09007}, \href{http://arxiv.org/abs/1405.7808}{{\tt
  arXiv:1405.7808}}\relax
\mciteBstWouldAddEndPuncttrue
\mciteSetBstMidEndSepPunct{\mcitedefaultmidpunct}
{\mcitedefaultendpunct}{\mcitedefaultseppunct}\relax
\EndOfBibitem
\bibitem{LHCb-DP-2013-003}
R.~Arink {\em et~al.}, \ifthenelse{\boolean{articletitles}}{\emph{{Performance
  of the LHCb Outer Tracker}},
  }{}\href{http://dx.doi.org/10.1088/1748-0221/9/01/P01002}{JINST \textbf{9}
  (2014) P01002}, \href{http://arxiv.org/abs/1311.3893}{{\tt
  arXiv:1311.3893}}\relax
\mciteBstWouldAddEndPuncttrue
\mciteSetBstMidEndSepPunct{\mcitedefaultmidpunct}
{\mcitedefaultendpunct}{\mcitedefaultseppunct}\relax
\EndOfBibitem
\bibitem{LHCb-DP-2012-002}
A.~A. Alves~Jr.\ {\em et~al.},
  \ifthenelse{\boolean{articletitles}}{\emph{{Performance of the LHCb muon
  system}}, }{}\href{http://dx.doi.org/10.1088/1748-0221/8/02/P02022}{JINST
  \textbf{8} (2013) P02022}, \href{http://arxiv.org/abs/1211.1346}{{\tt
  arXiv:1211.1346}}\relax
\mciteBstWouldAddEndPuncttrue
\mciteSetBstMidEndSepPunct{\mcitedefaultmidpunct}
{\mcitedefaultendpunct}{\mcitedefaultseppunct}\relax
\EndOfBibitem
\bibitem{LHCb-DP-2012-003}
M.~Adinolfi {\em et~al.},
  \ifthenelse{\boolean{articletitles}}{\emph{{Performance of the \lhcb RICH
  detector at the LHC}},
  }{}\href{http://dx.doi.org/10.1140/epjc/s10052-013-2431-9}{Eur.\ Phys.\ J.\
  \textbf{C73} (2013) 2431}, \href{http://arxiv.org/abs/1211.6759}{{\tt
  arXiv:1211.6759}}\relax
\mciteBstWouldAddEndPuncttrue
\mciteSetBstMidEndSepPunct{\mcitedefaultmidpunct}
{\mcitedefaultendpunct}{\mcitedefaultseppunct}\relax
\EndOfBibitem
\bibitem{LHCb-DP-2012-004}
R.~Aaij {\em et~al.}, \ifthenelse{\boolean{articletitles}}{\emph{{The \lhcb
  trigger and its performance in 2011}},
  }{}\href{http://dx.doi.org/10.1088/1748-0221/8/04/P04022}{JINST \textbf{8}
  (2013) P04022}, \href{http://arxiv.org/abs/1211.3055}{{\tt
  arXiv:1211.3055}}\relax
\mciteBstWouldAddEndPuncttrue
\mciteSetBstMidEndSepPunct{\mcitedefaultmidpunct}
{\mcitedefaultendpunct}{\mcitedefaultseppunct}\relax
\EndOfBibitem
\bibitem{spheno}
W.~Porod, \ifthenelse{\boolean{articletitles}}{\emph{{SPheno, a program for
  calculating supersymmetric spectra, SUSY particle decays and SUSY particle
  production at $e^+ e^-$ colliders}},
  }{}\href{http://dx.doi.org/10.1016/S0010-4655(03)00222-4}{Comput.\ Phys.\
  Commun.\  \textbf{153} (2003) 275},
  \href{http://arxiv.org/abs/hep-ph/0301101}{{\tt arXiv:hep-ph/0301101}}\relax
\mciteBstWouldAddEndPuncttrue
\mciteSetBstMidEndSepPunct{\mcitedefaultmidpunct}
{\mcitedefaultendpunct}{\mcitedefaultseppunct}\relax
\EndOfBibitem
\bibitem{SPS_benchmarks}
B.~C. Allanach, \ifthenelse{\boolean{articletitles}}{\emph{{The Snowmass points
  and slopes: benchmarks for SUSY searches}},
  }{}\href{http://dx.doi.org/10.1007/s10052-002-0949-3}{Eur.\ Phys.\ J.\
  \textbf{C25} (2002) 113}, \href{http://arxiv.org/abs/hep-ph/0202233v1}{{\tt
  arXiv:hep-ph/0202233v1}}\relax
\mciteBstWouldAddEndPuncttrue
\mciteSetBstMidEndSepPunct{\mcitedefaultmidpunct}
{\mcitedefaultendpunct}{\mcitedefaultseppunct}\relax
\EndOfBibitem
\bibitem{prospino_paper}
W.~Beenakker, R.~Hoepker, and M.~Spira,
  \ifthenelse{\boolean{articletitles}}{\emph{{PROSPINO: A program for the
  PROduction of Supersymmetric Particles In Next-to-leading Order QCD}},
  }{}\href{http://arxiv.org/abs/hep-ph/9611232}{{\tt
  arXiv:hep-ph/9611232}}\relax
\mciteBstWouldAddEndPuncttrue
\mciteSetBstMidEndSepPunct{\mcitedefaultmidpunct}
{\mcitedefaultendpunct}{\mcitedefaultseppunct}\relax
\EndOfBibitem
\bibitem{cteq66}
P.~M. Nadolsky {\em et~al.},
  \ifthenelse{\boolean{articletitles}}{\emph{{Implications of CTEQ global
  analysis for collider observables}},
  }{}\href{http://dx.doi.org/10.1103/PhysRevD.78.013004}{Phys.\ Rev.\
  \textbf{D78} (2008) 013004}, \href{http://arxiv.org/abs/0802.0007}{{\tt
  arXiv:0802.0007}}\relax
\mciteBstWouldAddEndPuncttrue
\mciteSetBstMidEndSepPunct{\mcitedefaultmidpunct}
{\mcitedefaultendpunct}{\mcitedefaultseppunct}\relax
\EndOfBibitem
\bibitem{prospino-errors}
M.~Kr{\"{a}}mer {\em et~al.},
  \ifthenelse{\boolean{articletitles}}{\emph{{Supersymmetry production cross
  sections in pp collisions at $\sqrt{s}$=7}},
  }{}\href{http://arxiv.org/abs/1206.2892}{{\tt arXiv:1206.2892}}\relax
\mciteBstWouldAddEndPuncttrue
\mciteSetBstMidEndSepPunct{\mcitedefaultmidpunct}
{\mcitedefaultendpunct}{\mcitedefaultseppunct}\relax
\EndOfBibitem
\bibitem{Sjostrand:2006za}
T.~Sj\"{o}strand, S.~Mrenna, and P.~Skands,
  \ifthenelse{\boolean{articletitles}}{\emph{{PYTHIA 6.4 physics and manual}},
  }{}\href{http://dx.doi.org/10.1088/1126-6708/2006/05/026}{JHEP \textbf{05}
  (2006) 026}, \href{http://arxiv.org/abs/hep-ph/0603175}{{\tt
  arXiv:hep-ph/0603175}}\relax
\mciteBstWouldAddEndPuncttrue
\mciteSetBstMidEndSepPunct{\mcitedefaultmidpunct}
{\mcitedefaultendpunct}{\mcitedefaultseppunct}\relax
\EndOfBibitem
\bibitem{Agostinelli:2002hh}
Geant4 collaboration, S.~Agostinelli {\em et~al.},
  \ifthenelse{\boolean{articletitles}}{\emph{{Geant4: A simulation toolkit}},
  }{}\href{http://dx.doi.org/10.1016/S0168-9002(03)01368-8}{Nucl.\ Instrum.\
  Meth.\  \textbf{A506} (2003) 250}\relax
\mciteBstWouldAddEndPuncttrue
\mciteSetBstMidEndSepPunct{\mcitedefaultmidpunct}
{\mcitedefaultendpunct}{\mcitedefaultseppunct}\relax
\EndOfBibitem
\bibitem{LHCb-PROC-2011-006}
M.~Clemencic {\em et~al.}, \ifthenelse{\boolean{articletitles}}{\emph{{The
  \lhcb simulation application, Gauss: Design, evolution and experience}},
  }{}\href{http://dx.doi.org/10.1088/1742-6596/331/3/032023}{{J.\ Phys.\ Conf.\
  Ser.\ } \textbf{331} (2011) 032023}\relax
\mciteBstWouldAddEndPuncttrue
\mciteSetBstMidEndSepPunct{\mcitedefaultmidpunct}
{\mcitedefaultendpunct}{\mcitedefaultseppunct}\relax
\EndOfBibitem
\bibitem{mstw08}
A.~D. Martin {\em et~al.}, \ifthenelse{\boolean{articletitles}}{\emph{{Parton
  distributions for the LHC}},
  }{}\href{http://dx.doi.org/10.1140/epjc/s10052-009-1072-5}{Eur.\ Phys.\ J.\
  \textbf{C63} (2009) 189}, \href{http://arxiv.org/abs/0901.0002}{{\tt
  arXiv:0901.0002}}\relax
\mciteBstWouldAddEndPuncttrue
\mciteSetBstMidEndSepPunct{\mcitedefaultmidpunct}
{\mcitedefaultendpunct}{\mcitedefaultseppunct}\relax
\EndOfBibitem
\bibitem{dynnlo}
S.~Catani {\em et~al.}, \ifthenelse{\boolean{articletitles}}{\emph{{Vector
  boson production at hadron colliders: A fully exclusive QCD calculation at
  next-to-next-to-leading order}},
  }{}\href{http://dx.doi.org/10.1103/PhysRevLett.103.082001}{{Phys.\ Rev.\
  Lett.\ } \textbf{103} (2009) 082001},
  \href{http://arxiv.org/abs/0903.2120}{{\tt arXiv:0903.2120}}\relax
\mciteBstWouldAddEndPuncttrue
\mciteSetBstMidEndSepPunct{\mcitedefaultmidpunct}
{\mcitedefaultendpunct}{\mcitedefaultseppunct}\relax
\EndOfBibitem
\bibitem{LHCb-PAPER-2012-008}
LHCb collaboration, R.~Aaij {\em et~al.},
  \ifthenelse{\boolean{articletitles}}{\emph{{Inclusive $W$ and $Z$ production
  in the forward region at $\sqrt{s}=7$ TeV}},
  }{}\href{http://dx.doi.org/10.1007/JHEP06(2012)058}{JHEP \textbf{06} (2012)
  058}, \href{http://arxiv.org/abs/1204.1620}{{\tt arXiv:1204.1620}}\relax
\mciteBstWouldAddEndPuncttrue
\mciteSetBstMidEndSepPunct{\mcitedefaultmidpunct}
{\mcitedefaultendpunct}{\mcitedefaultseppunct}\relax
\EndOfBibitem
\bibitem{LHCb-PAPER-2014-047}
LHCb collaboration, R.~Aaij {\em et~al.},
  \ifthenelse{\boolean{articletitles}}{\emph{{Precision luminosity measurements
  at LHCb}}, }{}\href{http://dx.doi.org/10.1088/1748-0221/9/12/P12005}{JINST
  \textbf{9} (2014) P12005}, \href{http://arxiv.org/abs/1410.0149}{{\tt
  arXiv:1410.0149}}\relax
\mciteBstWouldAddEndPuncttrue
\mciteSetBstMidEndSepPunct{\mcitedefaultmidpunct}
{\mcitedefaultendpunct}{\mcitedefaultseppunct}\relax
\EndOfBibitem
\bibitem{LHCb-PAPER-2015-001}
LHCb collaboration, R.~Aaij {\em et~al.},
  \ifthenelse{\boolean{articletitles}}{\emph{{Measurement of the forward $Z$
  boson cross-section in $pp$ collisions at $\sqrt{s}=7$ TeV}},
  }{}\href{http://arxiv.org/abs/1505.07024}{{\tt arXiv:1505.07024}}, {submitted
  to JHEP}\relax
\mciteBstWouldAddEndPuncttrue
\mciteSetBstMidEndSepPunct{\mcitedefaultmidpunct}
{\mcitedefaultendpunct}{\mcitedefaultseppunct}\relax
\EndOfBibitem
\bibitem{LHCb-PAPER-2014-033}
LHCb collaboration, R.~Aaij {\em et~al.},
  \ifthenelse{\boolean{articletitles}}{\emph{{Measurement of the forward $W$
  boson production cross-section in $pp$ collisions at $\sqrt{s}=7$ TeV}},
  }{}\href{http://dx.doi.org/10.1007/JHEP12(2014)079}{JHEP \textbf{12} (2014)
  079}, \href{http://arxiv.org/abs/1408.4354}{{\tt arXiv:1408.4354}}\relax
\mciteBstWouldAddEndPuncttrue
\mciteSetBstMidEndSepPunct{\mcitedefaultmidpunct}
{\mcitedefaultendpunct}{\mcitedefaultseppunct}\relax
\EndOfBibitem
\bibitem{F-C}
G.~J. Feldman and R.~D. Cousins, \ifthenelse{\boolean{articletitles}}{\emph{{A
  unified approach to the classical statistical analysis of small signals}},
  }{}\href{http://dx.doi.org/10.1103/PhysRevD.57.3873}{{Phys.\ Rev.\ }
  \textbf{D57} (1998) 3873}\relax
\mciteBstWouldAddEndPuncttrue
\mciteSetBstMidEndSepPunct{\mcitedefaultmidpunct}
{\mcitedefaultendpunct}{\mcitedefaultseppunct}\relax
\EndOfBibitem
\bibitem{pole}
J.~Conrad, \ifthenelse{\boolean{articletitles}}{\emph{{Including systematic
  uncertainties in confidence interval construction for Poisson statistics}},
  }{}\href{http://dx.doi.org/10.1103/PhysRevD.67.012002}{Phys.\ Rev.\
  \textbf{D67} (2003) 012002},
  \href{http://arxiv.org/abs/hep-ex/0202013v2}{{\tt
  arXiv:hep-ex/0202013v2}}\relax
\mciteBstWouldAddEndPuncttrue
\mciteSetBstMidEndSepPunct{\mcitedefaultmidpunct}
{\mcitedefaultendpunct}{\mcitedefaultseppunct}\relax
\EndOfBibitem
\end{mcitethebibliography}

\newpage

\centerline{\large\bf LHCb collaboration}
\begin{flushleft}
\small
R.~Aaij$^{38}$, 
B.~Adeva$^{37}$, 
M.~Adinolfi$^{46}$, 
A.~Affolder$^{52}$, 
Z.~Ajaltouni$^{5}$, 
S.~Akar$^{6}$, 
J.~Albrecht$^{9}$, 
F.~Alessio$^{38}$, 
M.~Alexander$^{51}$, 
S.~Ali$^{41}$
% ...

{\footnotesize \it
% ...
$ ^{5}$Clermont Universit\'{e}, Universit\'{e} Blaise Pascal, CNRS/IN2P3, LPC, Clermont-Ferrand, France\\
$ ^{6}$CPPM, Aix-Marseille Universit\'{e}, CNRS/IN2P3, Marseille, France\\
% ...
$ ^{9}$Fakult\"{a}t Physik, Technische Universit\"{a}t Dortmund, Dortmund, Germany\\
% ...
$ ^{37}$Universidad de Santiago de Compostela, Santiago de Compostela, Spain\\
$ ^{38}$European Organization for Nuclear Research (CERN), Geneva, Switzerland\\
% ...
$ ^{41}$Nikhef National Institute for Subatomic Physics, Amsterdam, The Netherlands\\
% ...
$ ^{46}$H.H. Wills Physics Laboratory, University of Bristol, Bristol, United Kingdom\\
% ...
$ ^{51}$School of Physics and Astronomy, University of Glasgow, Glasgow, United Kingdom\\
$ ^{52}$Oliver Lodge Laboratory, University of Liverpool, Liverpool, United Kingdom\\
% ...
}
\end{flushleft}

\end{document}